\documentclass[12pt,preprint]{aastex}

\usepackage{epstopdf}
\usepackage{natbib}
\usepackage{graphicx}
\usepackage{xcolor}
\usepackage {rotating, graphicx}
\newcommand{\tco}{\ifmmode {^{13}{\rm CO}} \else {$^{13}{\rm CO}$}\fi}
\newcommand{\dco}{\ifmmode {^{12}{\rm CO}} \else {$^{12}{\rm CO}$}\fi}
\newcommand{\juz}{\ifmmode {{\rm J}=1\rightarrow 0} \else
{J=1$\rightarrow$0}\fi}
\newcommand{\jdu}{\ifmmode {{\rm J}=2\rightarrow 1} \else
{J=2$\rightarrow$1}\fi}
\newcommand{\jtd}{\ifmmode {{\rm J}=3\rightarrow 2} \else
{J=3$\rightarrow$2} \fi}

\begin{document}

%% LaTeX will automatically break titles if they run longer than
%% one line. However, you may use \\ to force a line break if
%% you desire.

\title{A Multi-Scale Picture of Magnetic Field and Gravity from Large-Scale Filamentary Envelope to 
Core-Accreting Dust Lanes in the High-Mass Star-Forming Region W51}
%\author{Patrick M. Koch$^1$ et al.}
%\affil{$^{1}$Academia Sinica, Institute of Astronomy and Astrophysics, Taipei, Taiwan}

\author{Patrick M. Koch$^1$, Ya-Wen Tang$^1$, Paul T.P. Ho$^{1,2}$, Pei-Ying Hsieh$^{3,4}$, Jia-Wei Wang$^1$, 
Hsi-Wei Yen$^1$, 
Ana Duarte-Cabral$^5$, 
Nicolas Peretto$^5$, and 
Yu-Nung Su$^1$}

\affil{$^{1}$Academia Sinica, Institute of Astronomy and Astrophysics, Taipei, Taiwan}
\affil{$^{2}$East Asian Observatory (EAO), 660 N. Aohoku Place, University Park, Hilo, Hawaii 96720, USA}
\affil{$^{3}$Joint ALMA Observatory, Alonso de C\'ordova, 3107, Vitacura, Santiago 763-0355, Chile}
\affil{$^{4}$European Southern Observatory, Alonso de C\'ordova, 3107, Vitacura, Santiago 763-0355, Chile}
\affil{$^{5}$School of Physics \& Astronomy, Cardiff University, Queen's Building, The Parade, Cardiff, CF24 3AA, UK}

\email{pmkoch@asiaa.sinica.edu.tw}

\begin{abstract}
We present 230 GHz continuum polarization observations with the Atacama Large Milimeter/Submillimeter Array (ALMA)
at a resolution of 0$\farcs1$ ($\sim 540$~au) in the high-mass star-forming regions W51 e2 and e8. These observations 
resolve a network of core-connecting 
dust lanes, %filamentary extensions and fibers in dust continuum, 
marking a departure from earlier coarser 
more spherical continuum structures. At the same time, the cores do not appear to fragment further. 
Polarized dust emission is clearly detected.  The inferred magnetic field orientations are prevailingly 
parallel to 
dust lanes. %extensions and fibers. 
This key structural feature is analyzed together with the local gravitational vector field.
The direction of local gravity is found to typically align with dust lanes.
%extensions and fibers.
With these findings we derive a stability criterion that defines a maximum magnetic field strength that can be overcome
by an observed magnetic field-gravity configuration. Equivalently, this defines a minimum field strength that can stabilize 
dust lanes %extensions and fibers 
against a radial collapse. 
We find that the detected 
dust lanes %fibers and extensions 
in W51 e2 and e8 are stable, hence possibly making them
a fundamental component in the accretion onto central 
sources, providing support for massive star formation models without the need of large accretion disks.
When comparing to coarser resolutions, covering the scales of envelope, global, and local collapse, we find recurring similarities
in the magnetic field structures and their corresponding gravitational vector fields. These self-similar structures point at 
a multi-scale collapse-within-collapse scenario until finally the scale of core-accreting 
dust lanes %fibers   
is reached where gravity 
is entraining the magnetic field and aligning it with the 
dust lanes. %fibers and extensions. 
\end{abstract}

\section{Introduction}

%The role of the magnetic field in the star-formation process is a topic that has been gaining 
%growing attention both from theoretical and numerical sides as well as from observational perspectives. 
%need to introduce filamentary structures (Doris, Yasuo; BISTRO paper)

The formation and evolution of molecular clouds, the sites of high-mass star formation, are a complex interaction 
between gravity, turbulence, and magnetic fields, covering orders of magnitudes in physical length and density
%(e.g., \citet{hennebelle19, li14,crutcher12}).
\citep[e.g.][]{hennebelle19, li14,crutcher12}.
Moreover, a variety of feedback mechanisms add to the intricacy of the formation processes. 
Among all these constituents, the magnetic (B-)field still poses the likely biggest challenge to the establishment of 
a firm picture of star formation across time and scale.  
This has been largely due to the difficulty of detecting
signals that originate from the presence of B-fields (as they are typically only at the percent level of non-magnetic
signals) and the limited techniques to measure a B-field strength to gauge its significance against other constituents.
Recent advances in observational capabilities, offering substantially improved sensitivities,
are now rapidly changing this situation. 
Of particular interest are observations of dust continuum polarization. This is because 
a growing suite of instruments covering complete ranges in wavelengths and resolutions is available, 
and because dust polarization observations typically lead to the most connected and complete coverage in detections, unlike
e.g., Zeeman observations that remain challenging and are often limited to more localized and small areas in a source.

When utilizing dust polarization observations in the (sub-)millimeter regime, 
dust grains are thought to be aligned with their shorter axis parallel to the B-field. Rotating detected polarization 
orientations by $90^{\circ}$ then yields magnetic field orientations
\citep{cudlip82, hildebrand84, hildebrand88, lazarian00, andersson15}.
At the densities and scales probed with the here presented observations, radiative torques can provide an 
explanation for this B-field-dust alignment
\citep{draine96, draine97, lazarian00, cho05, lazarian07,hoang16}.
A growing literature is mapping magnetic field structures based on this property and investigating both statistical 
findings and detailed higher-resolution features.
The survey conducted by \citet{zhang14} with the SubMillimeter Array (SMA) towards a sample of 
14 high-mass star-forming regions 
resolving scales around 0.1~pc (resolutions $\theta$ of $1\arcsec$ to several arcseconds) at 345~GHz provides
statistical evidence for magnetic fields playing an important role during the collapse and fragmentation of
massive molecular clumps.
Further enlarging this sample to 18 massive dense cores, the recent work in \citet{palau21} finds a tentative
positive correlation between the number of fragments and the mass-to-flux ratio, hinting that magnetic fields
can possibly suppress fragmentation.  Investigating the role of the B-field in the infrared dark cloud G14.225$-$0.506
using the Caltech Submillimeter Observatory  (CSO) observations with SHARP with $\theta\sim 10\arcsec$
at 350$\mu$m, \citet{anez21} find that different B-field morphologies and strengths can explain the different observed
fragmentation properties. Also observed with the CSO/SHARP, the different fragmentation types in 
G34.43$+$0.24 are explained by a different relative significance of gravity, turbulence, and magnetic field \citep{tang19}.
Observed with POL-2 on the James Clerk Maxwell Telescope (JCMT) with $\theta\sim 14\arcsec$ at 850 $\mu$m, 
the NGC 6634 filamentary network 
is resolved down to about 0.1~pc, revealing detailed B-field structures and variations across the main filament and 
sub-filaments \citep{arzoumanian21}. 
Higher-resolution (sub-)arcsecond observations with ALMA (mostly around 850 $\mu$m and 1.2~mm) have started
to reveal detailed morphological features in the magnetic field, such as an expanding UCHII region in G5.89$-$0.39
leaving a clear imprint in the B-field morphology \citep{fernandez21}, sharpening the earlier coarser 
SMA observations \citep{tang09a}; a highly fragmented filament in W43-MM1 \citep{cortes16}; 
a resolved hour-glass magnetic field structure in G31.41$+$0.31 \citep{beltran19}; and
ring-like and arm-like structures likely resulting from toroidal wrapping of the magnetic field in OMC-3 \citep{takahashi19}.

W51 is a high-mass star-forming complex at parallax distances around 5.41~kpc for W51 e2 and e8 \citep{sato10}
and 5.10~kpc for W51 North \citep{xu09}, located in a region with little foreground
and background contamination.
The entire complex shows star-formation activities at various evolutionary stages \citep{ginsburg17, saral17, ginsburg15}.
Collimated small-scale SiO outflows are detected in W51e2-E, e8, and North \citep{goddi20}, 
and they appear to connect to larger-scale outflows seen in $^{12}$CO(2--1) in all three sources \citep{ginsburg17} and also in 
$^{12}$CO(3--2) in e2-E \citep{shi10a}.
The plane-of-sky B-field morphology has been mapped with a series of polarization observations with
increasingly higher angular resolutions $\theta$, starting from the earliest interferometric observations with 
BIMA \citep[$\theta\sim3\arcsec$;][]{lai01}
to the SMA \citep[$\theta\sim0\farcs7$;][]{tang09b, tang13} and 
to the first observations with ALMA \citep[$\theta\sim0\farcs26$;][]{koch18}.
The BIMA observations at 1.3~mm showed W51 e2 and e8 as an elongated connected north-south structure with 
a magnetic field mostly perpendicular to it \citep{lai01}. The e2 region manifested itself as a clear polarization
hole. The higher-resolution SMA observations at 0.87~mm
%specifically targeting e2 and e8
revealed more 
complex magnetic field structures which are likely the reason for the depolarization in the larger BIMA beam. 
The finer B-field structures in e2 and e8 showing hourglass-like topologies with clearly bent field lines were interpreted as  
gravitational collapse imprinted onto the B-field morphology \citep{tang09b}. The first ALMA
observations at 1.3~mm
\citep{koch18}, again improving the resolution by a factor of 10 in area, revealed striking new features. 
In particular, they clearly resolved the satellite core e2-NW with bow-shock shaped B-field structures that
are hinting infall of this smaller core towards the dominating mass center e2-E. Additionally, areas with centrally 
converging symmetrical B-field structures (convergence zones)
and possibly streamlined B-field morphologies were detected. A generic 
feature seen in many of the resolved cores inside e2, e8, North, and also on larger scale between e2 and e8,
is B-field structures resembling gravitational pull
towards the core's center on one side with the other side showing B-field lines appearing to be dragged away
towards the next more massive neighboring core. This imprint in the B-field morphology was interpreted as a 
scenario where local collapse is ongoing while a locally collapsing core, as an entity, is pulled to the next more
massive gravitational center which itself is also collapsing \citep{koch18}.  
Recent numerical 
work by \citet{vazquez19} is exactly presenting such a scenario as a result of a global hierarchical collapse 
where a flow regime leads to collapses within collapses.

While the successively higher-resolution observations in W51 keep revealing new magnetic field features
from imprints of dynamical processes, the W51 region has, at the same time, served as a mine of information for our 
developments of new analysis techniques.  The SMA observations \citep{tang09b} served as a testbed 
for the polarization--intensity gradient technique \citep{koch12a, koch12b}. This technique uses the measurable 
angle $\delta$ between a magnetic field orientation and an intensity gradient as a key observable which, in 
combination with a second angle between intensity gradient and local gravity, makes it possible to derive 
a magnetic field strength. The technique gives a {\it local} magnetic field strength -- at every position where
a magnetic field orientation is detected -- and therefore, leads to {\it maps} of field strengths. 
At the same time, the technique puts forward a magnetic field-to-gravity force ratio, $\Sigma_{\rm{B}}$, 
based solely on measurable 
angles which allows for a completely independent estimate of a mass-to-flux ratio \citep{koch12b}. 
The establishment of $\delta$ as a prime observable as well as an approximation for $\Sigma_{\rm{B}}$
is presented in \citet{koch13} with an application to a 50-source sample of low- and high-mass star-forming
sources in \citet{koch14}. A main result from this series of papers is the 
{\it recognition of a spatially varying role of the magnetic field,}
e.g., mass-to-flux ratios can transition from outer sub-critical to inner 
super-critical areas in a star-forming region, and force ratios $\Sigma_{\rm{B}}$ are clearly varying from 
zones where collapse and infall are slowed down or prohibited by the magnetic field to other zones, 
within the same source, where collapse is possible. With the first ALMA data in \citet{koch18} 
an additional measure was introduced, the $\sin\omega$ measure. The angle $\omega$,
in the range between 0 and $90^{\circ}$, measures the projection of the local magnetic field tension force
along the local direction of gravity, and hence quantifies the fraction (in a range of 0 to 1) of the magnetic field tension
force that can work against the gravitational pull. Maps of $\sin\omega$ of all the cores in W51 e2, e8, and North 
systematically display zones where the magnetic field is maximally opposing gravity and other zones where 
the magnetic field is nearly or completely ineffective in slowing down gravity \citep{koch18}. 
It should be noted that all of these techniques are utilizing a combination of the geometry and shape of both 
the magnetic field and the underlying emission (density) structures to infer the local role of the magnetic field. 

As presented in the following sections, this current work is resolving once more finer structures
with a resolution of $\theta\sim0\farcs1$, an improvement in area by a factor of 7
over the earlier $0\farcs26$ observations,
reaching a physical length
scale of about 2.6~mpc or 540 au at the distance of W51 e2/e8. 
%The current work is resolving 
With this, the earlier near-spherical structures are resolved, 
%resolving the outer diffuse medium and 
revealing connecting dust lanes. %extensions and fiber-type connections. 
%It is with these data, 
%possible with ALMA's unprecedented sensitivity and resolution, 
%that the pursuit of the role of the magnetic field towards the next finer structures is possible. 
%Here, we are presenting the latest high-resolution ALMA polarization observations ($\theta\sim0\farcs1$)
%resolving a physical linear scale of xy mpc.  
Together with earlier observations covering larger scales we propose a 
synergetic multi-scale scenario of the evolving role of the magnetic field in the W51 high-mass star-forming region starting from 
the large filamentary envelope scale ($\sim 0.5$~pc), global-collapsing-core scale ($\sim 0.05$~pc), inside-core fragmenting scale
($\sim 10$~mpc) down to the scale of dust lanes
%finger-like accreting fiber scale 
($\sim 2.6$~mpc) accreting onto central cores.
The paper is organized as follows.
Section \ref{section_observations} describes our ALMA observations. Polarization properties are given in 
the appendix. The detected key structural features are 
introduced in Section \ref{key_features}. 
Section \ref{section_analysis} analyzes the gravitational vector field together with 
the magnetic field morphology and derives a stability criterion for filaments and fibers. 
The discussion in Section \ref{section_discussion} presents a road towards a synergetic multi-scale picture.

%%%%%%%%%%%%%%%%%%

\section{Observations}
\label{section_observations}

The project was observed with the ALMA Band 6 receiver
(around a wavelength of 1.3~mm)
in Cycle 4 and Cycle 5, project codes
\#2016.1.01484.S and \#2017.1.01242.S.
Observations were done in two execution blocks (EBs) on August 17, 2017.
The two EBs were calibrated separately in flux, bandpass, and gain.
The polarization calibrations were performed after merging the two calibrated EBs. 
The array included 44 antennas with (projected) baselines ranging from 21 m to 3638 m.
The four basebands were set in TDM mode (64 channels for a 2 GHz bandwidth per baseband).
The calibration (bandpass, phase, amplitude, flux) was performed using CASA\footnote{http:\/\/casa.nrao.edu\/} v4.7.2. 
J1922+1530 (flux $\sim$ 0.219 Jy at 232.9 GHz) was used as a phase calibrator, and 
J1751+0939 and J1922+1530 were the flux calibrators.
%The polarization calibrator is J1924-2914, and the derived polarization ratio is 2.56\%, and the polarization angle is 45.6$\degr$, in agreement with other ALMA measurements.
The phase centers for W51 e2 and e8 where 
(RA, DEC)$=$(19:23:43.95, $+$14:30:34.00)  and 
(RA, DEC)$=$(19:23:43.90, $+$14:30:27.00), respectively, 
in J2000 coordinates.
The presented images are with a Briggs weighting scheme and a robust parameter of 0.5, 
which gives an angular resolution 
of 0$\farcs$11$\times$0$\farcs$10, with a position angle PA of -23$\degr$. 
The sensitivities of the Stokes I, Q, U images are 1.2 mJy/beam, 0.04 mJy/beam, and 0.04 mJy/beam, respectively.
Polarization measurements $I_p=\sqrt{Q^2+U^2}>0$ are positively biased (while both $Q$ and $U$ can
be negative).
Hence, $I_p$ in the high signal-to-noise regime ($I_p\geq3\sigma_p$) is debiased as  
$I_p=\sqrt{Q^2+U^2-\sigma_{Q,U}^2}$, where $\sigma_Q\approx \sigma_U$ are the noise levels in $Q$
and $U$ \citep{leahy89, wardle74}.
Investigating instrumental polarization in Band 6, \citet{nagai16} conclude that linear polarization at a level of $<0.1$\% is detectable. 
For images presented in this paper, the two simultaneous conditions of having Stokes  $I\geq$3$\sigma$ 
and $I_p\geq$3$\sigma_p$ are imposed.
Polarization averages are around 3\%, with a single minimum value of about 0.1\% and some isolated maximum values above 10\%. 
Detailed polarization results are in the appendix and in Figures \ref{figure_polarization} and \ref{figure_hist_pol_perc}.

\section{Observed Key Structural Features} \label{key_features}

In the following we are describing three new structural features in dust continuum Stokes $I$ and in dust polarization (B-field), 
seen in the $0\farcs1$ ($\sim 2.6$ ~mpc or $\sim 540$~au) resolution ALMA observations of W51 e2 and e8: 
(1) resolved networks of dust lanes in Stokes $I$; 
%extensions and fibers in Stokes $I$; 
(2) magnetic field morphology in dust lanes;
%extensions, fibers, and the transition to extensions; 
(3) sectors of straight field lines around e2-E followed by an abrupt change in magnetic field orientations in the central core region. 

\begin{sidewaysfigure}
\includegraphics[width=22cm]{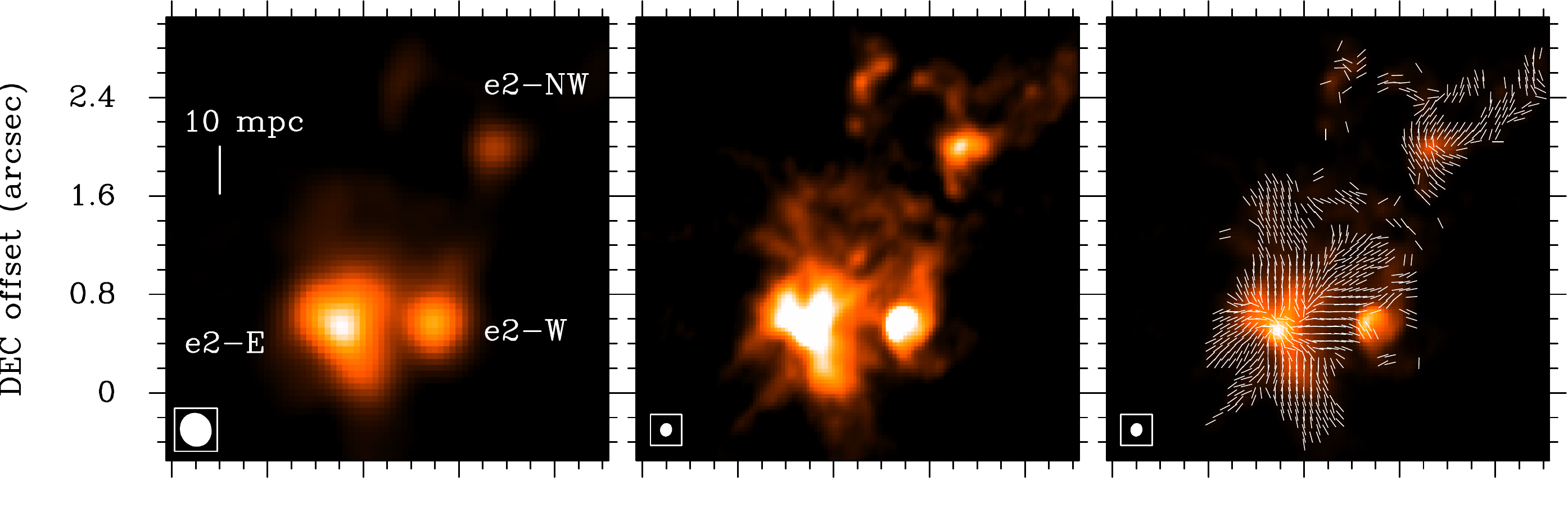}
\includegraphics[width=22cm]{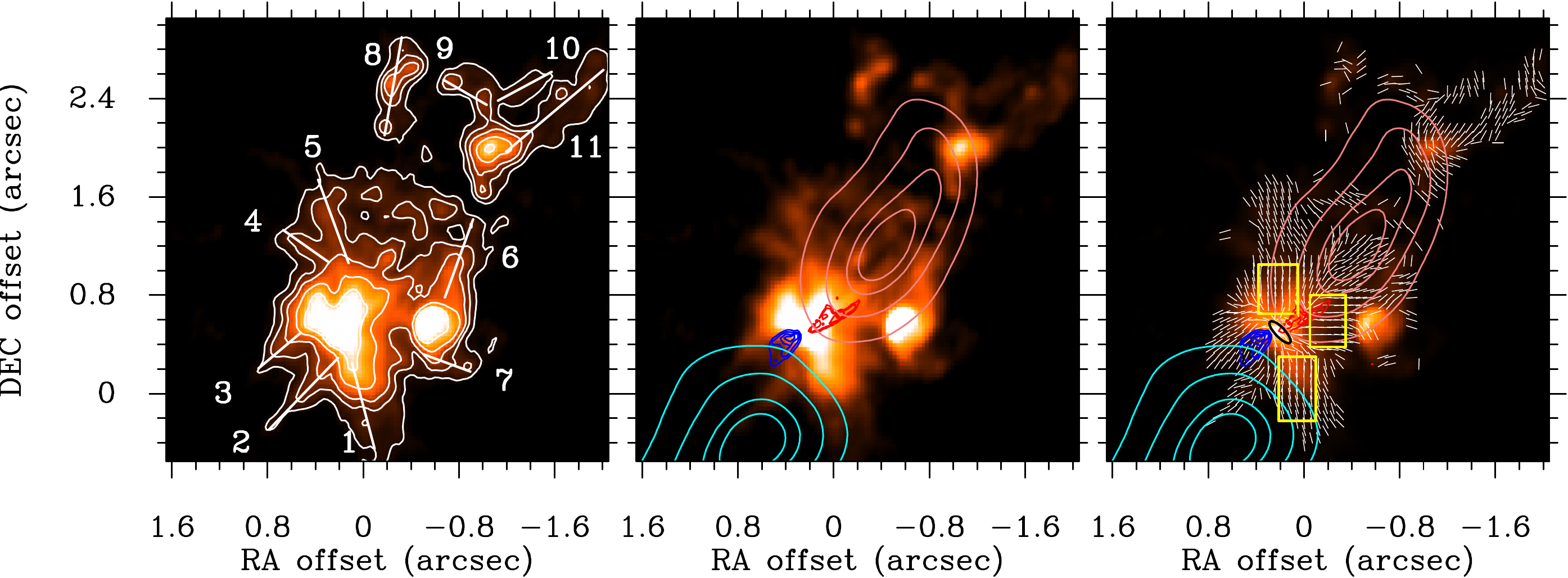}
\caption{\small Dust continuum emission in color scale with a resolution of 0\farcs1 ($\sim$2.6 mpc or $\sim$ 540 au) around 230 GHz ($\sim1.3$~mm) in Band 6 with ALMA for W51 e2, except the upper left panel which shows the earlier still more roundish detection with a 0\farcs26 resolution \citep{koch18}.
White segments display the magnetic field in the right panels at half of the synthesized beam resolution (white ellipses in the lower left corners). 
Outflows are over-plotted in the bottom middle and right panel. 
The cyan and faint-red contours denote the blue- and red-shifted $^{12}$CO 2-1 outflow component, and the blue and red contours mark the blue- and red-shifted SiO 5-4 outflow emission.
Dust lanes and streamers are labelled in the bottom left panel where continuum contours are identical 
to the ones described in the right panels in Figure \ref{figure_synergy_1}. 
 The bottom right panel marks the sectors of nearly straight field lines (yellow rectangles) and the possible small disk (black ellipse) as discussed in section \ref{key_features}.}
\label{figure_B}
\end{sidewaysfigure}

\begin{sidewaysfigure}
\includegraphics[width=22cm]{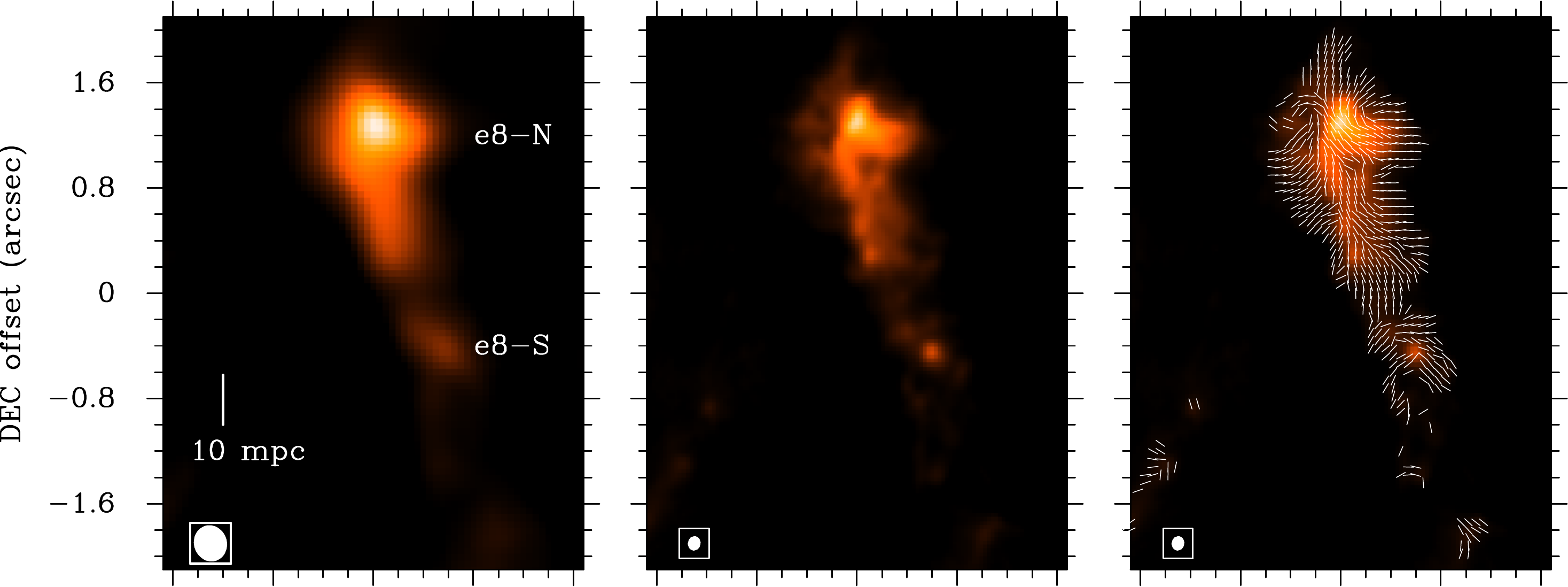}
\includegraphics[width=22cm]{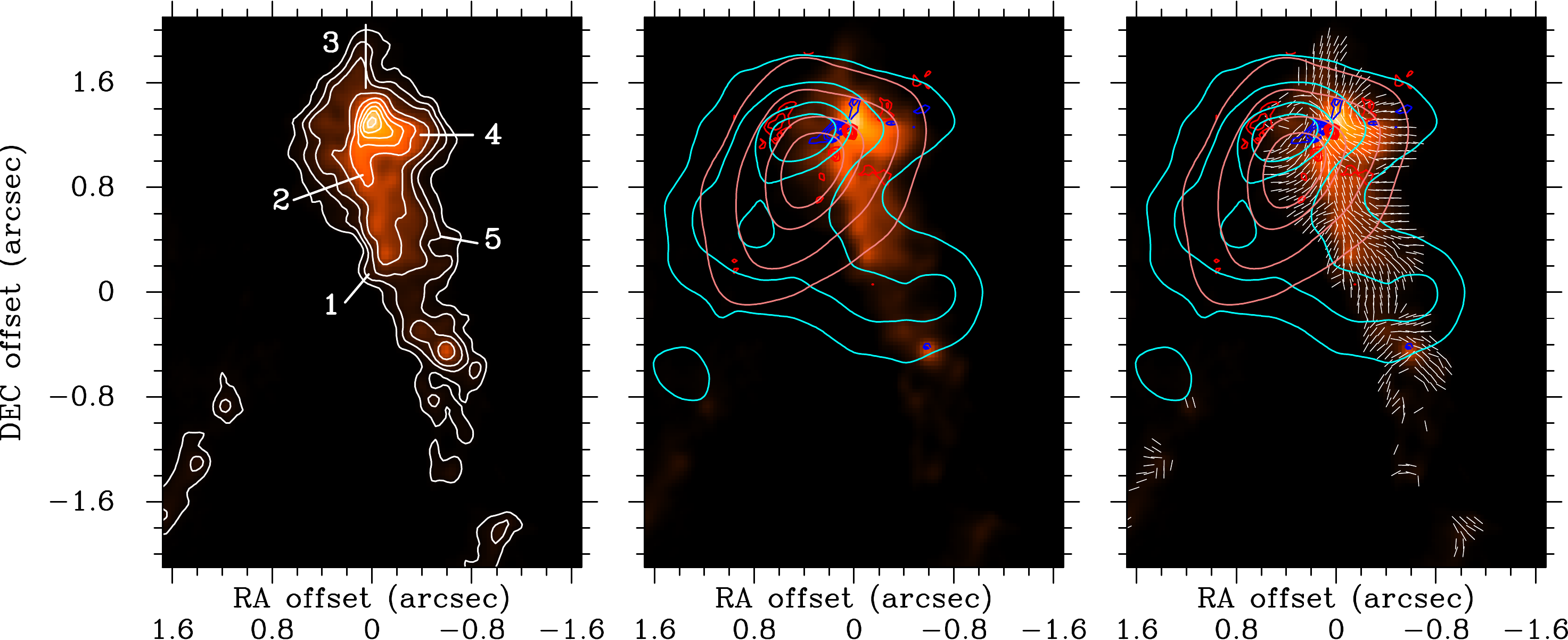}
\caption{Identical to Figure \ref{figure_B} but for W51 e8.}
\label{figure_B_2} 
\end{sidewaysfigure}

{\it Departure from spherical continuum structures -- resolved networks of core-connecting dust lanes.}
Earlier SMA \citep[$\theta\sim 3\arcsec-0\farcs7$;][]{tang09b} 
and ALMA \citep[$0\farcs26$;][]{koch18} observations showed
dust continuum structures that appeared circular (e2) and smoothly elongated (e8), with no indications yet of 
resolved shapes, sizes, and their surroundings. Improving the resolution 
from the ALMA $0\farcs26$ to our latest ALMA $0\farcs1$ observations -- giving a finer resolution of about a factor
of 7 in area -- starts to reveal a radically different 
picture (Figure \ref{figure_B} and \ref{figure_B_2}).
The cores e2-E, e2-W, e2-NW, e8-N, and
e8-S do not seem to fragment further. The previously circular dust continuum emission around these cores is
resolved into streamer-like lanes that appear to connect to the central and more compact cores. 
All these lanes appear to either come in from the cores' peripheries or they form connections 
between individual cores. 
%In the following we will refer to these extensions also as 'fibers'. 
For e2, we start to 
resolve a network of dust lanes, converging towards e2-E and e2-W.  The satellite core e2-NW is further 
resolved into three tail-like streamers (feature 9, 10, and 11, as labelled in the bottom left panel in Figure \ref{figure_B})
where two are along a NW-SE direction.  A more isolated lane in the north (lane 8), west 
of e2-NW, is pointing to the main core e2-E.  Around the dominating agglomerate e2-E/e2-W we find 
at least seven dust lanes (lane 1 to 7), all roughly radially pointing to the central agglomerate. They vary 
in their resolved widths from about $0\farcs1$ to $0\farcs4$ ($\sim 540$~au to $2,000$~au) 
with lengths around $0\farcs5$ before
merging into the central denser cores.  W51 e8 (Figure \ref{figure_B_2}, bottom left panel)
displays
a main north-south connection between e8-S and e8-N with a width and length of about $0\farcs5$ and 1\arcsec, respectively.
The e8-N core
shows three main emerging lanes from its surrounding, namely from the south, north, and west (lanes 2, 3, and 4).
Two possible lanes, hinted by outward bulking contours (lanes 1 and 5), are merging into the main north-south connection.
%where the south lane connects to e8-S.  
Though being evident in their emerging shape, these extensions and dust lanes are not yet as clear as in W51 e2. 
Section \ref{analysis_local_gravity} with Figure \ref{figure_loc_grav} and \ref{figure_grav_stream} motivates 
and explains these dust lanes based on the underlying gravitational structures.

{\it Magnetic field in dust lanes.}
%extensions and fibers, and transition from surrounding diffuse region to denser extensions.} 
With the resolved dust continuum structures, the plane-of-sky projected magnetic field morphology is 
also resolved in the dust lanes.
%extensions and fibers. 
%Without any exceptions, 
The B-field orientations 
are prevailingly aligned parallel along the dust lanes.
%to the orientations of the extensions and fibers. 
In W51 e2, the B-field 
is observed to be parallel in the three dust lanes 1 to 3 in the south converging to e2-E 
(Figure \ref{figure_B}, bottom left and top right panel), 
and also in the lane in the north (lane 5). 
Almost no polarization is detected along the northeastern dust lane (lane 4).
%Here, the polarization is 
%not detected in the thinner outer beginning of the {\bf lane}, but around the offset (R.A., Dec.)$\sim(0.25, 1.5)$,  
%where the {\bf lane}
%broadens, the B-field appears also parallel. 
The e2-E and e2-W cores appear connected with straight 
field lines.  No or only very incomplete polarization is detected in the 
emerging lane 7
%two wider extensions 
in the west of e2-W. 
The longer tail-like streamer of e2-NW (lane 11) shows a B-field morphology clearly aligned with the tail. 
The thinner northern tail (lane 10) has an initial field structure that appears perpendicular to the tail in its northern
tip while then becoming more aligned moving along the tail to the south. 
The dust lane 8 is disconnected and displays varying field orientations.
Since W51 e8 is not yet as resolved as e2, our observations seem to additionally capture the transition in 
the B-field structures from the surrounding diffuse material to the (relatively denser) emerging dust lanes
(Figure \ref{figure_B_2}, bottom left and top right panel). 
The region between e8-S and e8-N illustrates this.  The thin bridge immediately north of e8-S (around 
offset (R.A., Dec.)$\sim(-0.25, -0.1)$) already shows a B-field along a north-south axis.
Moving further north, the B-field on the lower
contour levels on the east and west side is mostly perpendicular to while progressively bending and 
getting aligned with the e8-N -- e8-S axis in the inner relatively denser spine region of the main north-south connection. 
This feature is very similar to the findings in W51 e2 with the coarser resolution of $0\farcs26$ \citep{koch18}
where field lines symmetrically converging from two sides towards a central line were observed. This feature
was named a {\it convergence zone}, and it was interpreted as material being accreted symmetrically from 
two sides towards a central region or channel.  Since e8 appears less resolved, we might be seeing this same 
feature as already seen in e2 with coarser resolution. Hence, this transition from outer perpendicular lines to 
the inner more aligned field lines might be a more common feature.  
The dust lanes around e8-N (lanes 2, 3, and 4)
have their field lines aligned and pointing towards the center of e8-N. 
The possible lanes 1 and 5 show field line orientations that follow the general trend of the convergence zone.

{\it Straight magnetic field lines in three sectors around W51 e2-E and abrupt change in field orientations in the 
innermost center region.}
The magnetic field morphology shows nearly straight plane-of-sky projected field lines in 
three sectors around the Stokes $I$ emission peak of W51 e2-E 
(Figure \ref{figure_B}, bottom right panel).
%These nearly straight field lines 
%are observed in the western quadrant towards e2-W, the northern quadrant around offset (R.A., Dec.)$\sim(0.25, 0.25)$, 
%and the southern quadrant around offset (R.A., Dec.)$\sim(0, 0)$. 
All three sectors extend over an area of about $0\farcs5\times0\farcs5$. 
The field lines within these sectors appear to be 
at 90$\degr$ angles with respect to each other (from North, to West, to South), 
with field orientations along a north-south direction (northern and southern sector), and along an 
east-west direction (western sector).
%The quadrant 
%east of e2-E displays still an overall east-west field orientation, though not as strikingly regular
%as the other three quadrants. 
An overall trend of converging {\it bent} field lines towards e2-E is already seen in the coarser
$0\farcs26$ resolution image in \citet{koch18}. The higher resolution maps here show strikingly
{\it straight} field lines in three sectors around e2-E. 
%{\it Abrupt change in magnetic field orientations in the innermost center regions in e2 and e8.}
Following the three sectors closer to the emission peak of e2-E, offset
(R.A., Dec.)$\sim(0.2, 0.5)$, the nearly straight B-field lines are abruptly changing orientations, forming 
a prevailingly northeast-southwest aligned structure. This structure is aligned roughly 45$\degr$
with respect to the north-south and east-west field orientations in the three sectors.
%the three above identified quadrants. 
The change in orientation from the nearly straight outer field lines
to this inner structure happens within one or two synthesized beams, with changes between 
adjacent field segments as large as almost 90$\degr$. While the field morphologies in the 
three sectors and the centermost region are all coherent and connected, they are also clearly 
different. This likely indicates that these two regions are governed by two different physical processes. 
We note that both the blue- and red-shifted SiO outflow lobe seem to originate from where the 
magnetic field lines transition into the 45$\degr$-oriented central field pattern 
(Figure \ref{figure_B}, bottom right panel).
As this outflow is interpreted as a signpost of an underlying not-yet-resolved disk \citep{goddi20}, we speculate that this innermost structure is capturing the B-field morphology towards and in this putative
accretion disk.
In W51 e8, the e8-N core shows close-to-straight incoming field lines towards its peak from the 
west, north and from the south (Figure \ref{figure_B_2}, top right panel).
Although not as regular as in e2-E, rapid changes in field orientations
around 90$\degr$ are also observed towards the peak location,
where again an accretion disk is expected based on a detected SiO outflow
\citep[Figure \ref{figure_B_2}, bottom right panel;][]{goddi20}.

\section{Analysis} 
\label{section_analysis}

\subsection{Direction of Local Gravity around Cores and in Dust Lanes} \label{analysis_local_gravity}

\begin{figure}
\includegraphics[width=9cm]{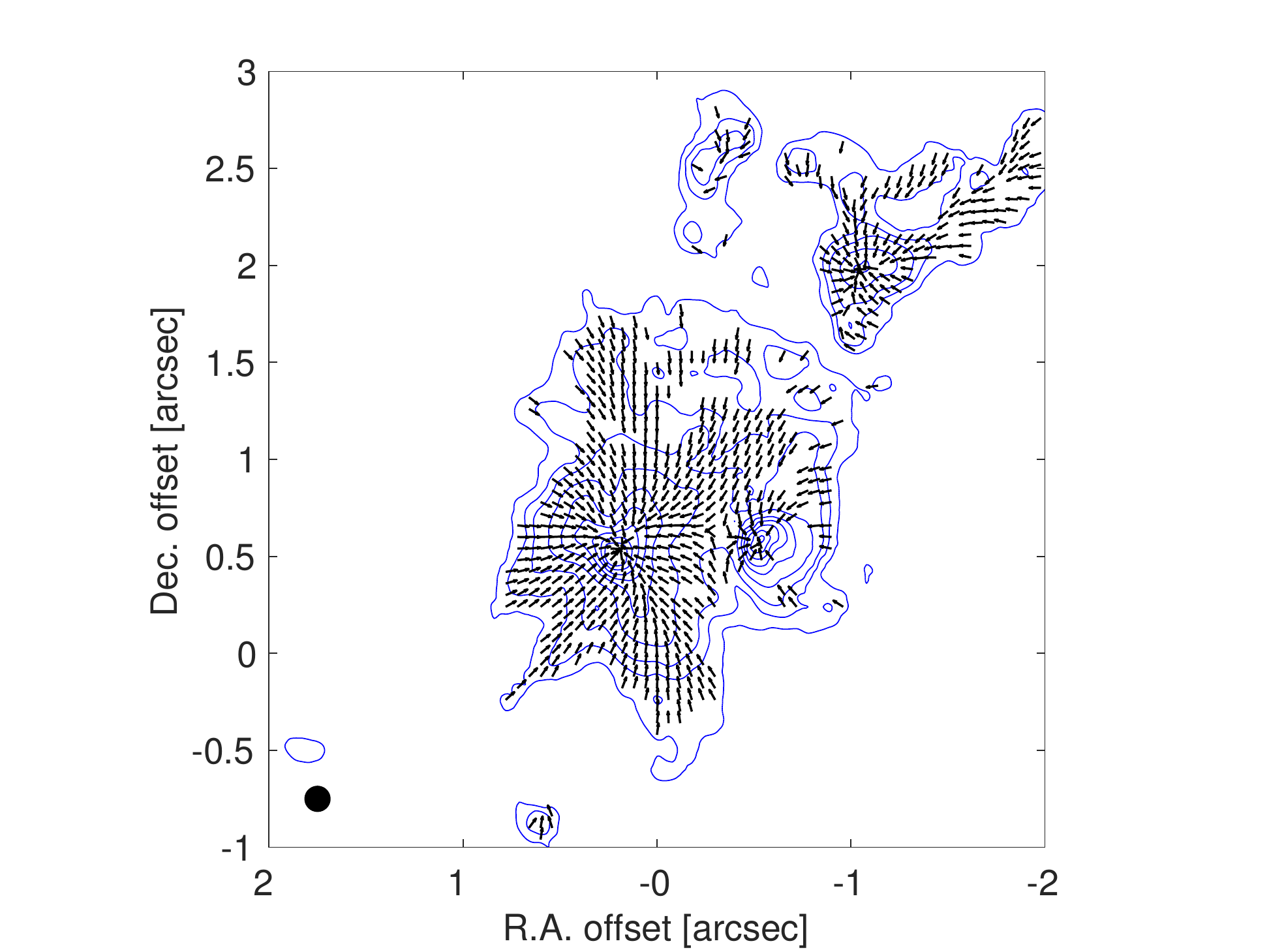}
\includegraphics[width=9cm]{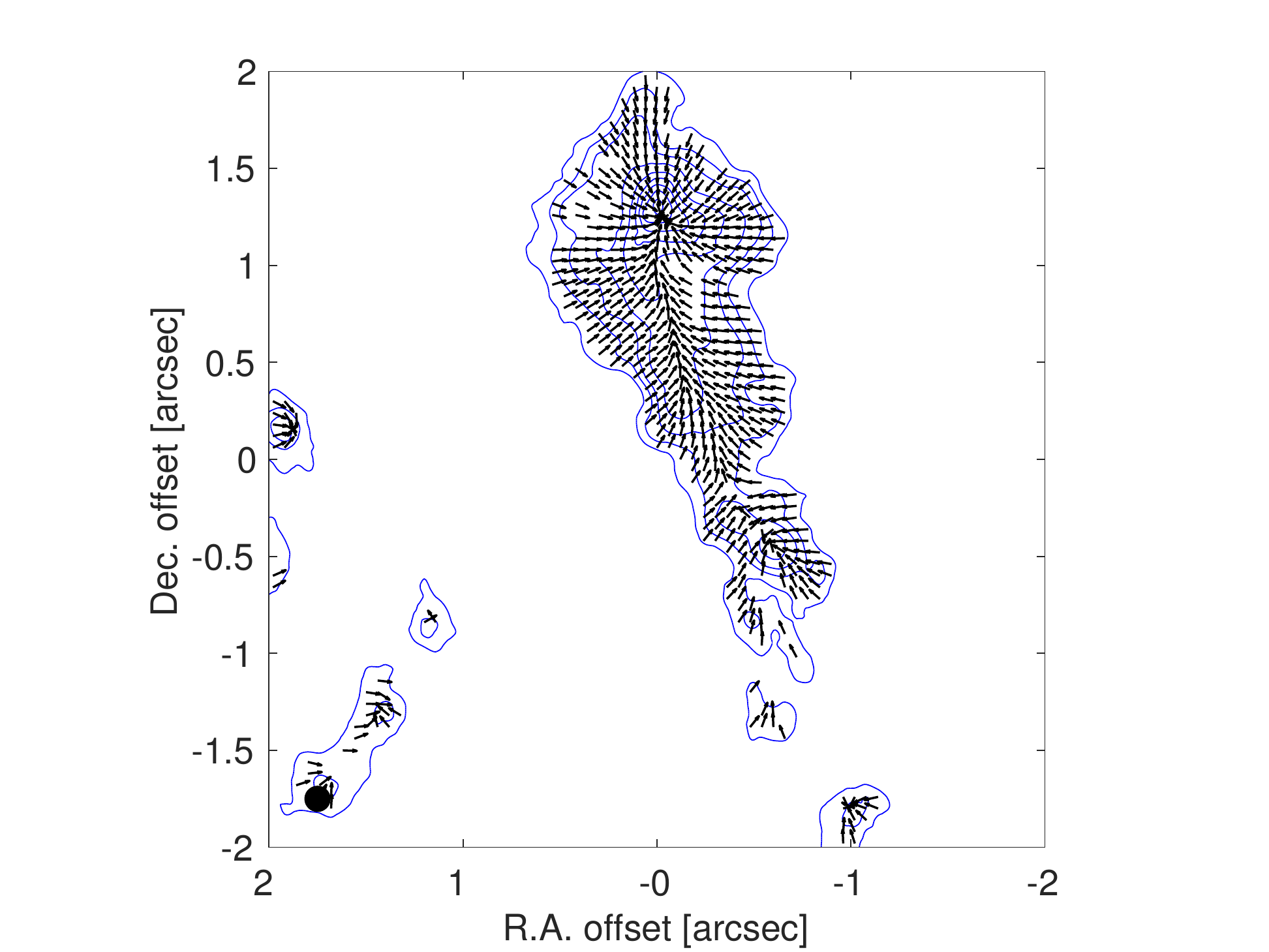}
\caption{Local gravitational vector fields $\mathbf{g}_{\rm{loc}}$ for W51 e2 (left panel) and e8 (right panel) 
overlaid on dust continuum contours (blue). 
Vectors are displayed with a uniform length, giving the direction but not the 
absolute magnitude of local gravity.  Vectors are selectively shown for the locations where polarized emission is detected and 
displayed at half of the synthesized beam resolution.}
\label{figure_loc_grav} 
\end{figure}

How is gravity acting in the connecting dust lanes in W51 e2 and e8?
Calculating the {\it local direction of gravity} in the plane of sky was introduced in 
\citet{koch12a,koch12b}. 
In the following we are interested in identifying systematic patterns in the 
gravitational vector field, both in conjunction with the underlying Stokes $I$
dust continuum structures and the observed B-field morphology. 
Figure \ref{figure_loc_grav} presents maps of the directions of local gravity for e2 and e8. 

A plane-of-sky projected direction and a relative magnitude of local gravity (at every location in a map)
can be derived by summing up all the surrounding pixelized dust continuum emission. This
sum is a vector sum that takes into account the directions and distances to every
surrounding pixel weighted by the dust emission. This vector sum is calculated at every pixel 
\citep{koch12a,koch12b}.
In order to have a measure for how gravity is locally acting, we need to assume 
that the distribution of dust emission is a fair tracer for the distribution of the total mass in a source. 
If this is the case, we can visualize the direction of the local gravitational pull (Figure \ref{figure_loc_grav}). 
The absolute magnitude of the gravitational pull is scalable by a gas-to-dust mass
ratio.  We note that an unknown (constant) gas-to-dust
mass ratio does not change the directions of the local gravity vectors. This ratio 
is only needed in order to go from relative to absolute magnitudes. For the following analysis
and discussion we assume that the observed dust distribution is closely enough tracing 
the overall mass distribution in W51 e2 and e8. 
In order to generate the gravitational vector fields in Figure \ref{figure_loc_grav}, an upper and lower limit
of distances needs to be defined within which the vector sum is calculated. 
The smallest scale is defined by the (synthesized) beam resolution, and the largest scale is given
by the largest distance to any detected emission in our maps. As we have tested, diffuse
emission at distances beyond the extension of our maps can be safely omitted. This is because
this emission is down-weighted by a quickly dropping $1/r^2$ factor and, furthermore,
the already diffuse emission tends to become more azimuthally symmetrical at larger distances
which will lead to largely canceling out any gravitational pull.

W51 e2 reveals an overall radial vector field, with a main converging point towards e2-E. 
While vectors in the south, north, and west of e2-W are still pointing towards the emission
peak of e2-W,  the vectors in the eastern end of e2-W (around offset (R.A., Dec.)$\sim(-0.5, 0.6)$) are already influenced
by the locally dominating mass associated with e2-E. As a result, these vectors 
are already being deflected and turning towards e2-E. 
The satellite core e2-NW shows a pattern where local gravity vectors from the north, northwest, and northeast
are converging around offset (R.A., Dec.)$\sim(-1.0, 2.0)$.
% south of the emission peak of e2-NW. 
It is interesting to note that the extension from the peak of e2-NW towards south, until offset (R.A., Dec.)$\sim(-1.0, 1.6)$, 
shows initial local gravity directions towards this converging point with a growing trend for directions
turning away towards southwest (the dominating mass concentration around e2-E) the further one moves 
along this extension to the south. 
This trend of {\it observing a local gravity field that is converging towards a local peak emission on one side while
the other side shows gravity vectors being bent away to a neighboring more dominating mass concentration} was
already noted in the coarser resolution data of W51 e2, e8, and North in \citet{koch18}.  The present observations
are further sharpening this picture, and they provide additional support for a {\it local collapse scenario}, where 
collapse can happen locally around this emission peaks while the underlying forming cores, as an entity, are being
pulled to the neighboring larger mass concentration (see also discussion in section \ref{discussion_multi_scale_collapse}).

\begin{figure}
\includegraphics[width=9cm]{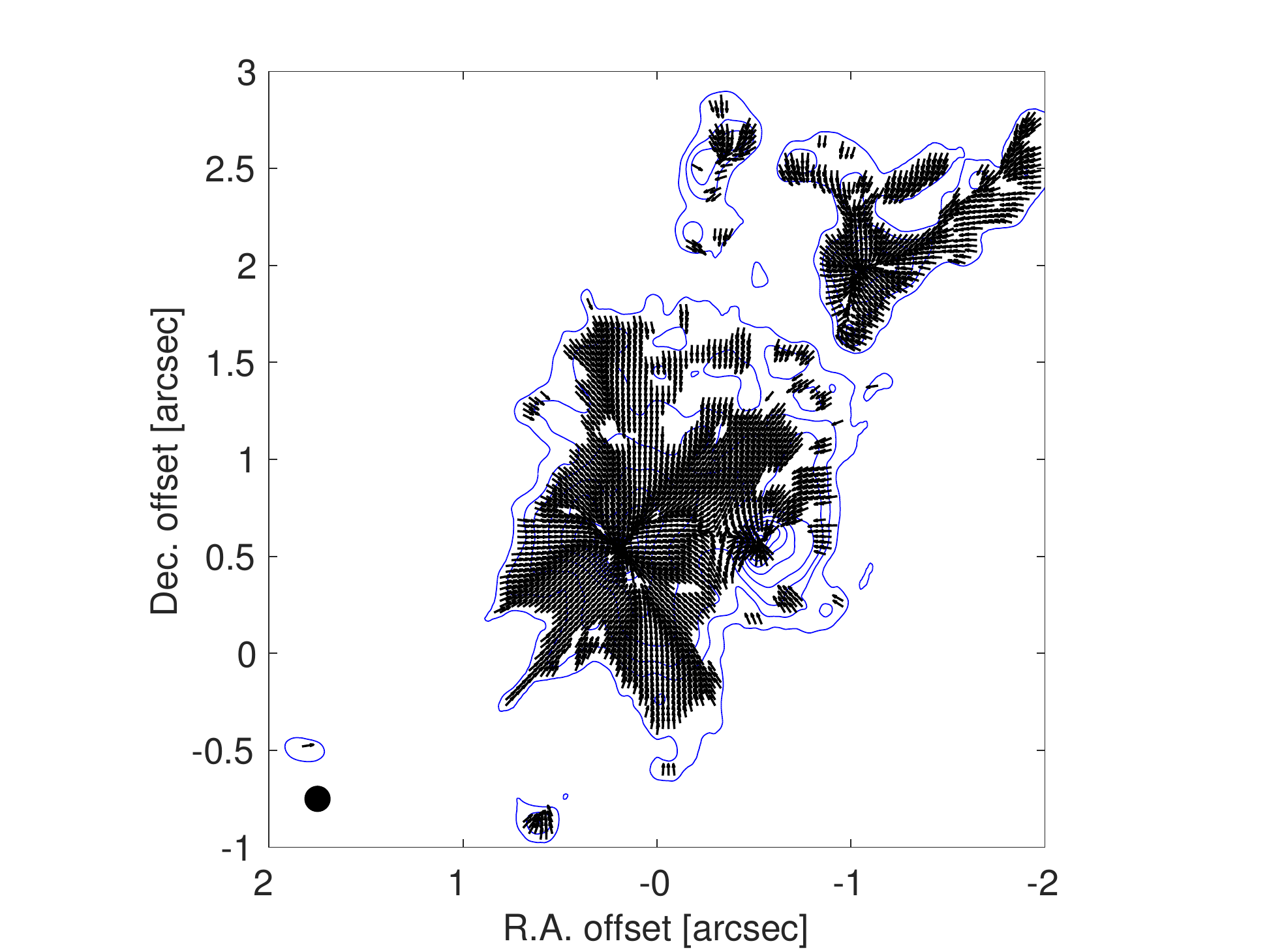}
\includegraphics[width=9cm]{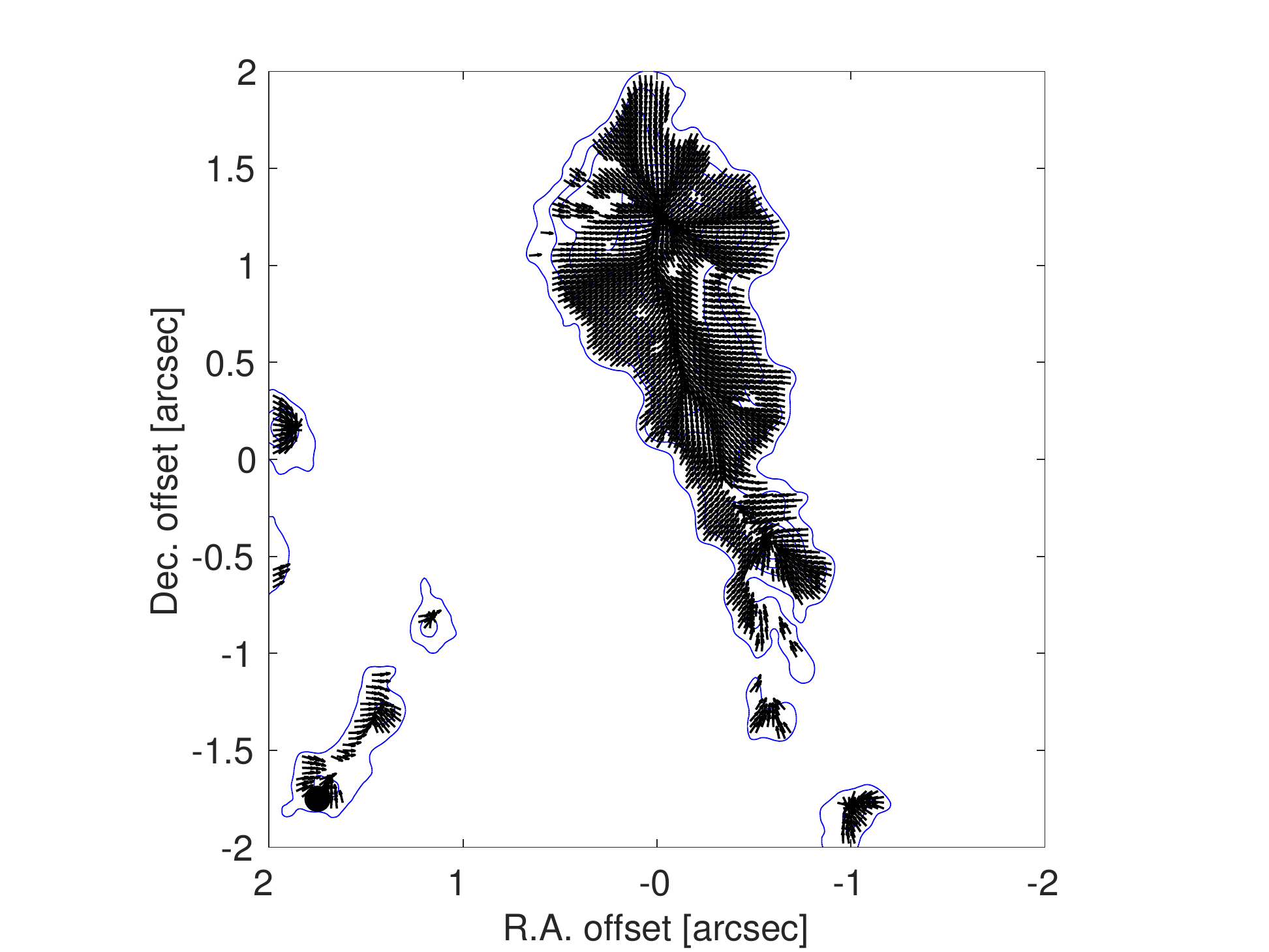}
\caption{
Identical to Figure \ref{figure_loc_grav} but with the gravitational vector field $\mathbf{g}_{\rm{loc}}$ oversampled to emphasize the gravitational local convergence towards ridges, before merging into local mass centers.
These ridges form a gravitational skeleton (visible as thicker black zones) and are suggestive of gravitationally 
driven accretion regions and channels. This skeleton largely falls onto the dust lanes labelled in 
Figure \ref{figure_B} and \ref{figure_B_2}.
}
\label{figure_grav_stream} 
\end{figure}

The local gravity field in all the dust lanes around e2-E/W 
(lanes 1 to 7, bottom left panel in Figure \ref{figure_B})
is overall radial. When overgridded for a better visualization, it additionally shows convergence towards
ridges that form a gravitational skeleton (Figure \ref{figure_grav_stream}) and likely are
the underlying structures for the dust lanes seen in Figure \ref{figure_B} and \ref{figure_B_2}.
As such, local gravity is 
dominantly directed {\it along} dust lanes. The two lanes coming in from the NW connecting
to e2-NW (lane 10 and 11) show a vector field mostly along their longer axes towards the e2-NW emission peak.
%, with the more northern extension showing
%less alignment but a direction towards the e2-NW emission peak.  
The exceptions of this overall alignment 
between dust lanes and local gravity directions are the southern extension of e2-NW, 
around offset (R.A., Dec.)$\sim(-1.0, 1.7)$, and dust lane 9.
%and the faint eastern dust lane, around offset (R.A., Dec.)$\sim(-0.7, 2.5)$. 
As discussed above for the 
southern extension, they both reveal a gravitational vector field being directed to the dominating southern
e2-E/e2-W agglomerate.

We remark that both converging gravitational centers, around e2-E and e2-NW, show an offset
with respect to their dust emission peaks. This offset is about $0\farcs15$ for e2-E along a northwest direction (towards e2-NW), 
and it is about $0\farcs25$ for e2-NW along a southeast direction (towards e2-E). 
While this might be due to e.g., projection effects or other forces that shape the dust distribution, 
the noted offsets are also close to our resolution.
The precise reason for these possible offsets is being further investigated. 

W51 e8 shows an overall radial vector field with a main converging point on e8-N around (R.A., Dec.)$\sim(0.0, 1.2)$.
Along its southern extension, this source displays a gravitational field that is converging to a north-south
oriented ridge along R.A. $\sim 0.0$. From east and west the vector field is first streaming into this central ridge
in the south before turning to become gradually more radial when approaching the e8-N core further north. 
The overall direction is pointing to the dominating e8-N core as far south as (R.A., Dec.)$\sim(-0.2, -0.1)$.
The e8-S core is a local gravitational center around (R.A., Dec.)$\sim(-0.5, -0.5)$ with a clear converging field 
from the southwest. Moving further north along a southwest-northeast axis this vector field is gradually turning
around, redirected, and aligning with the central ridge.

In summary, both e2 and e8 appear to have a single main gravitational center (e2-E and e8-N) with an overall simple
radial-like gravitational vector field. 
When overgridding the gravitational field, finer structures become more visible pointing at radial ridges forming
a gravitational skeleton (Figure \ref{figure_grav_stream}). This is likely the underlying structure that is defining and 
shaping the dust lanes, possibly together with feedback mechanisms as discussed in section \ref{section_applicability}.
Both e2 and e8 also show additional local gravitational centers (e2-W and e2-NW; e8-S)
that display a characteristic pattern with vectors pointing towards the center on one side and vectors being gradually 
deflected towards the main gravitational center on the other side.

%\subsection{Close overall resemblance local gravity vs B}
%
%???needed ???
%overall orientations very similar (sin omega maps?, numbers?)
%especially: east of e2-W and also south of satellite e2-NW; similar trends, maybe delay (-> refer to upcoming paper
%about oscillator; generic trend)

\subsection{B-field-induced Stability along Dust Lanes}
\label{analysis_B_induced_stability}

We further investigate the consequences of the two observational findings, namely (1) the B-field being prevailingly
aligned with dust lanes (section \ref{key_features}), 
and (2) the direction of local gravity being typically aligned
with dust lanes (section \ref{analysis_local_gravity}). 
A conclusion of these two findings is that local B-field orientations and the directions of local gravity 
show a very close overall resemblance. They are preferentially aligned with each other 
in the plane of sky, meaning that gravitational pull is mostly acting along B-field lines. 
%We note that 3D versus 2D geometry [COMPLETE; Tomisaka 2015; BISTRO NGC 1333]. 
Such an arrangement favors gas motions along magnetic field lines. On top of the already intrinsic preference
of motions along field lines rather than across field lines, a gravitational pull selectively acting parallel to the field 
lines is additionally supporting this. This makes motions along the radial direction in (dust) lanes
significantly more difficult.  As a consequence, any bulk infall or local collapse along the radial direction in a (dust) lane
or extension would need to both overcome the magnetic field tension force and counteract a (dominating)
gravitational pull along these structures. 
Given this constellation, we are asking the questions: {\it Is further collapse along a radial direction 
in these (dust) lanes and extensions
possible at all with the observed geometry and structure? Can these (dust) lanes and extensions fragment further, or 
are they the feeding and accreting network that connects to the central cores?}

In the following we are estimating the stabilizing role of the magnetic field in a scenario of possible further 
fragmentation and local collapse.

\subsubsection{Net Local Gravity versus B-field Tension}

\begin{figure}
\includegraphics[width=15cm]{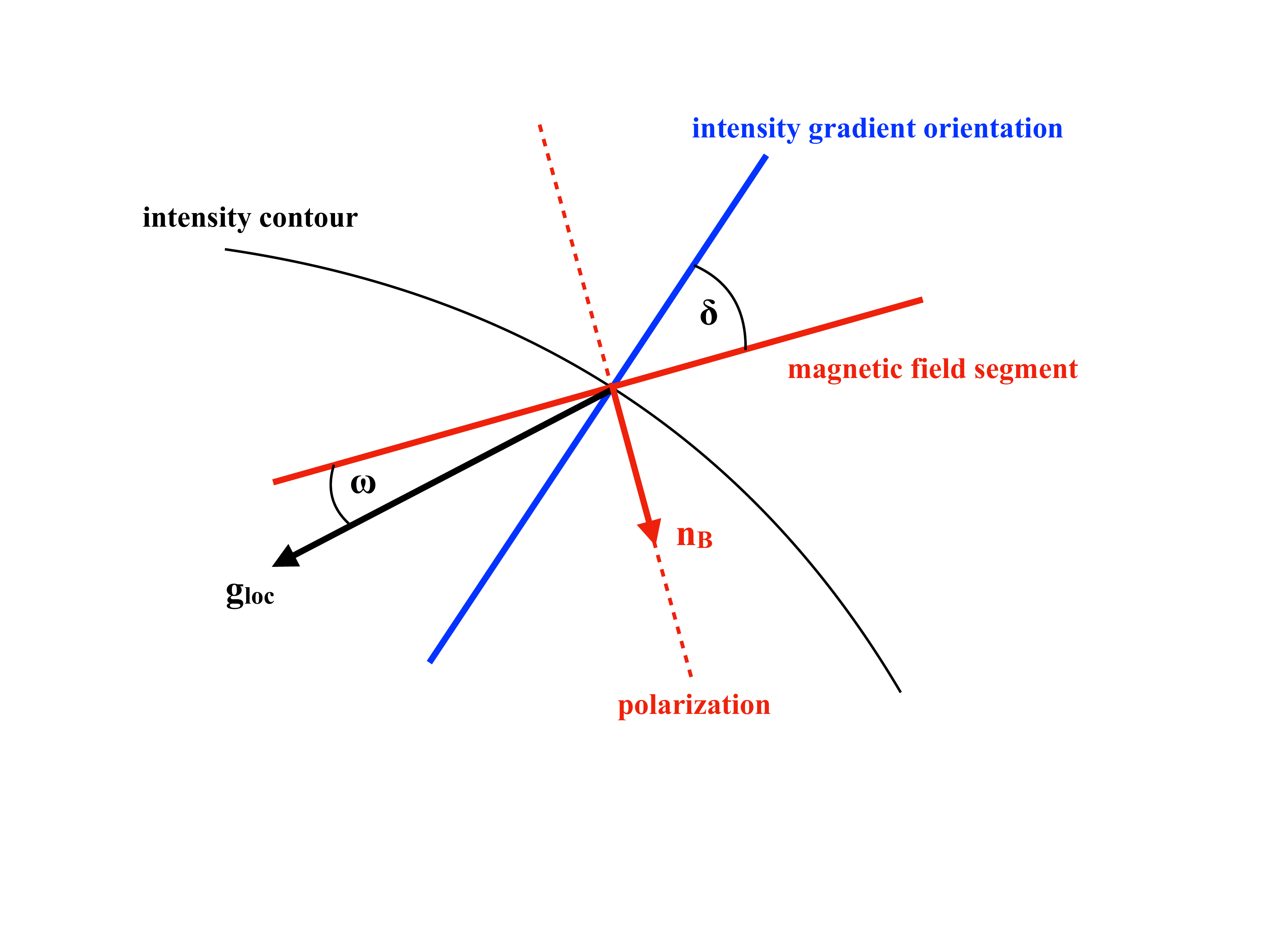}
\caption{Illustration of relevant angles. $\omega$ measures the deviation between the magnetic field orientation (in the case of submm dust polarization emission rotated by 90$^{\circ}$ from the originally detected polarization orientation) and the direction of local gravity $\mathbf{g}_{\rm{loc}}$. 
%$\mathbf{n}_g$ is orthogonal to $\mathbf{g}$ and forms an orthonormal system with it.
$\mathbf{n}_B$ is the unity vector perpendicular to a measured B-field orientation along the direction of the field tension force. 
$\delta$ is the angle between a magnetic field orientation and the intensity gradient orientation. 
%Figure is reproduced from \citet{koch18}. 
}
\label{figure_schematic_omega} 
\end{figure}

In order to initiate a local collapse, local gravity needs to overcome the magnetic field force and start to bend the 
field lines. Figure \ref{figure_schematic_omega}  illustrates the geometry adopted from earlier papers \citep{koch12a, koch18}.
{\it How much gravitational force is available locally to bend and drag a B-field line?} This is measured by the projection of the 
local gravity force $\mathbf{g}_{\rm{loc}}$ onto the direction of the (restoring) field tension $\mathbf{n}_{\rm{B}}$, 
orthogonal to a B-field line:
\begin{equation}
F_{\rm{loc,c}} \equiv\cos\left(\frac{\pi}{2}-\omega\right) \cdot |\mathbf{g}_{\rm{loc}}|= \sin\omega \cdot|\mathbf{g}_{\rm{loc}}|, 
                                                                                                                           \label{eq_f_loc}
\end{equation}
where $F_{\rm{loc, c}}$ is introduced as the net local force (after projection) that can trigger a collapse.
Hence, the observed misalignment $\omega (\le \pi/2)$ between a magnetic field orientation and the local gravity direction quantifies
the fraction of the local gravity force that can work to overcome the magnetic field. This fraction is in the range between 0 
and 1. The local gravity is directed along a field line with no force component at all to overcome the B-field  
if $\omega=0$, leading to $F_{\rm{loc,c}} = 0$, and the local gravity is maximally working against the B-field 
if $\omega=\pi/2$ (i.e., local gravity force $\mathbf{g}_{\rm{loc}}$
orthogonal to a B-field orientation, resulting in $F_{\rm{loc,c}} =|\mathbf{g}_{\rm{loc}}|$). 

We note that the initial motivation for Equation (\ref{eq_f_loc}) is different from the $\sin\omega$-measure introduced in 
\citet{koch18}. The above equation is motivated by assessing how much of an existing local gravitational pull is effectively directed
towards bending and dragging a B-field line, i.e., the local gravitational pull is projected onto the direction of the field tension.
In the case of the $\sin\omega$-measure, the B-field tension force is projected onto the local gravity direction in order 
to measure what fraction of the B-field tension force can impede a gravity-driven motion. In both cases, $\sin\omega$ is 
quantifying a fraction between 0 and 1, but it is the {\it fraction of the local gravitational pull} in the above Equation (\ref{eq_f_loc})
and the {\it fraction of the B-field tension force} in \citet{koch18}.

\subsubsection{Stability and Collapse Criterion}

\begin{figure}
\includegraphics[width=15cm]{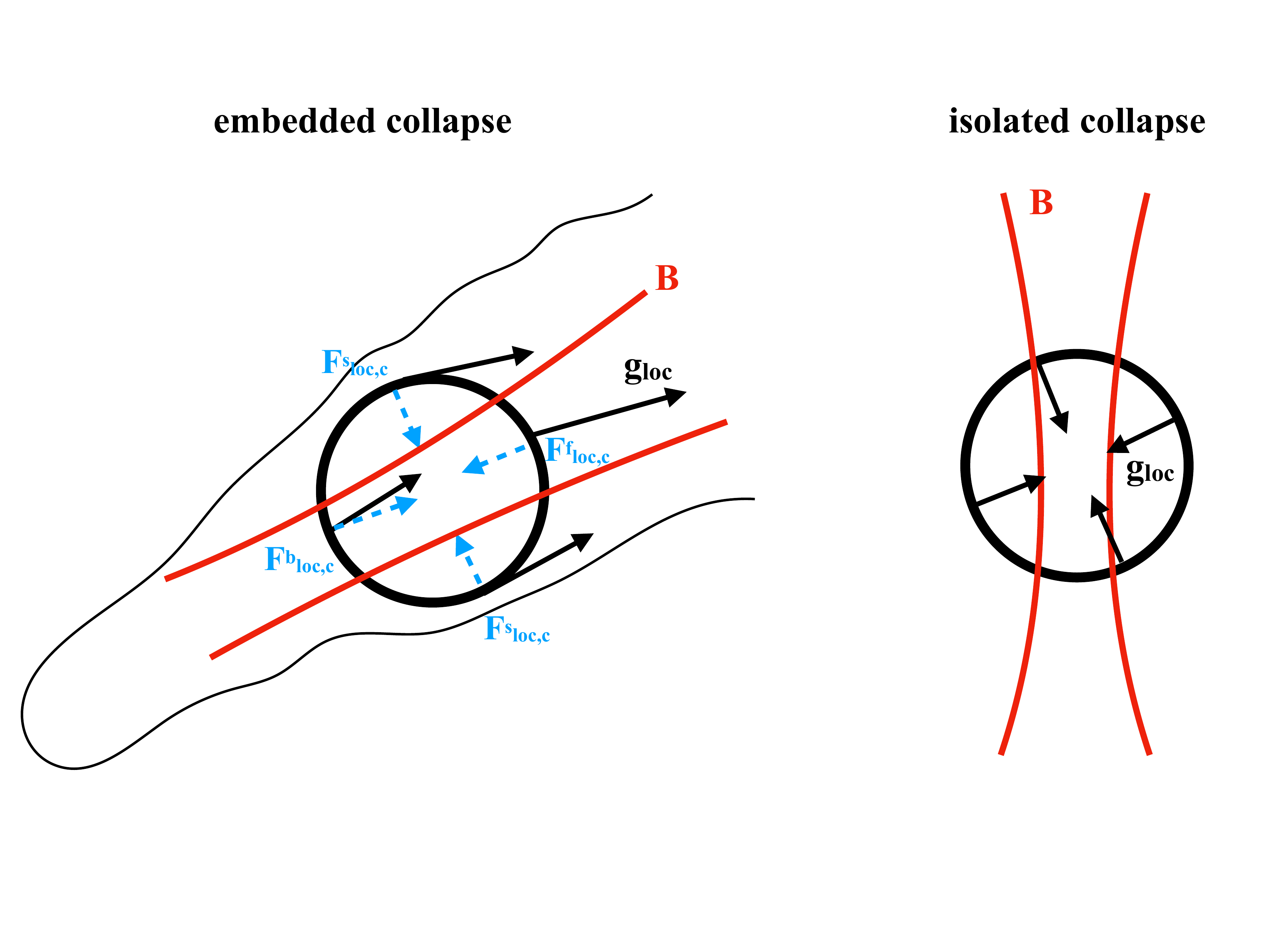}
\caption{Illustration of the difference between an isolated collapse (right side) and a collapse embedded in 
a dust lane (left side).  
The black circles symbolize spherical test volumes to initiate a collapse in the 
%structures undergoing collapse in the 
presence of a magnetic field (red lines). 
The local gravitational field, $\mathbf{g}_{\rm{loc}}$, is shown with black arrows.
In the case of the isolated collapse, this vector field is pointing towards the center of the sphere, and the orientation 
of the magnetic field with respect to this vector field is irrelevant because $\mathbf{g}_{\rm{loc}}$ is azimuthally symmetric. 
For the embedded collapse, $\mathbf{g}_{\rm{loc}}$ is the result of the surrounding mass distribution. Illustrated is 
a typical situation as observed in Figure \ref{figure_loc_grav}, 
where the local gravitational field along a filamentary dust lane is
directed towards the dominating center of mass (in this case towards the right side).  Generally, the individual 
vectors of $\mathbf{g}_{\rm{loc}}$ vary in length (i.e., strength of gravitational pull) and direction. 
The gravitational pull necessary for an embedded collapse, $F_{\rm{loc, c}}$,
is shown with the blue dashed vectors along 4 symbolic
directions (front, upper index 'f'; back, upper index 'b'; side, upper index 's').  Unlike for the isolated case, the relative 
orientation between field lines and $\mathbf{g}_{\rm{loc}}$ is pivotal because $\mathbf{g}_{\rm{loc}}$ is not 
azimuthally symmetric. As a consequence, the fraction and magnitude of $\mathbf{g}_{\rm{loc}}$ that is directed towards 
overcoming the field tension depends on this relative orientation. In particular, for the illustrated configuration, 
the net available gravitational pull from the sides, $F_{\rm{loc,c}}^{\rm{s}}$ is very small, 
if $\mathbf{g}_{\rm{loc}}$ and the magnetic field are prevailingly collinear. }
\label{figure_schematic_isolated_vs_filament} 
\end{figure}

From the observed magnetic field morphologies and the gravitational vector field in Figure \ref{figure_loc_grav}  
it is clear 
that the plane-of-sky projected B-field and local gravity orientations are well aligned in many areas along 
the dust lanes.
If fragmentation and subsequent local collapse are to occur in such a constellation, a local volume needs to become 
supercritical and decouple from the larger-scale ambient environment in an elongated 
dust lane.  A collapse
criterion for an {\it isolated} spherically symmetric volume, given by the requirement that the gravitational force
dominate over the B-field tension force, $F_{\rm{grav}}>F_{\rm{tension}}$, was given by \citet{schleuning98}, approximating
$F_{\rm{tension}}=1/(4\pi)(\mathbf{B}\cdot \nabla)\mathbf{B}$ as $1/(4\pi) B^2/R$ 
with the magnetic field curvature $1/R$ where $R$ is the field radius.
In this hypothetical case of an isolated sphere, the gravitational force field is spherically symmetric and pointing at the center of the sphere. 
Because of this symmetry,  the orientation of an initially uniform or hourglass-like B-field with respect to the spherical volume has no influence on a later collapse.
In particular, the local gravity force $\mathbf{g}_{\rm{loc}}$, which can enable collapse, is azimuthally symmetric in this 
isolated case. This is unlike our observed local field-gravity constellation where $\mathbf{g}_{\rm{loc}}$ has a clear 
directional dependence and where the magnetic field orientation with respect to the orientation of 
a dust lane and a possible subsequent collapse makes a decisive difference. 
Figure \ref{figure_schematic_isolated_vs_filament} 
schematically illustrates 
the difference between an isolated collapse and a collapse embedded in a dust lane.

For our observed constellation we can derive a collapse criterion as follows. 
Along a dust lane we are making the simplifying assumption 
$F_{\rm{loc,c}}^{\rm{f}}\approx F_{\rm{loc,c}}^{\rm{b}}\approx F_{\rm{loc,c}}$, 
i.e.,  the gravitational pull in the front, $F_{\rm{loc,c}}^{\rm{f}}$, is equal to the gravitational pull in the back,
$F_{\rm{loc,c}}^{\rm{b}}$. In reality, the pull in the front
might be slightly larger due to the smaller $1/r^2$ distance term to the dominating centre of gravity 
(e.g., e2-E in the case of e2). 
In any case, the directions of both $F_{\rm{loc,c}}^{\rm{f}}$ and $F_{\rm{loc,c}}^{\rm{b}}$ are pointing towards
the main centre of gravity which is unlike in the case of the isolated sphere where local gravity from two opposite sides around the 
sphere is pointing towards the sphere's centre (Figure \ref{figure_schematic_isolated_vs_filament}). 
The observed constellation clearly seems
to favor gas movements along the magnetic field lines along a single direction but not converging to a local center from two opposite 
sides. The slight difference between $F_{\rm{loc,c}}^{\rm{f}}$ and $F_{\rm{loc,c}}^{\rm{b}}$
might further mean that gas is actually accelerated differentially and stretched, all towards the main gravitating center. 
This leads to a first conclusion that local gravitational collapse inside a dust lane or extension, if really to happen, will need 
substantial compression from the side provided by $F_{\rm{loc,c}}^{\rm{s}}$. But this compression from two opposite 
sides has to overcome a main obstacle, namely the magnetic field tension force (equation (\ref{eq_f_loc})). 
Given the prevailingly close alignment between local gravity and local magnetic field orientation
($\omega$ small), we have
$F_{\rm{loc,c}}^{\rm{s}}=\cos(\pi/2-\omega)|\mathbf{g}_{\rm{loc}}| \ll  |\mathbf{g}_{\rm{loc}}|$.
Hence, a collapse in an extension or dust lane can only be enabled via a radial infall or collapse orthogonal to the local field line. 
With this and combining with the above expression for the field tension $F_{\rm{tension}}$ by \citet{schleuning98}, we
can write the criterion $F_{\rm{tension}}<F_{\rm{loc,c}}$ which yields
\begin{equation}
B < \sqrt{4\pi R \cdot |\mathbf{g}_{\rm{loc}}| \cdot \sin\omega}.    \label{eq_collapse_criterion}
\end{equation}

Adopting Gaussian-base units where $1\rm{G} = 1\rm{g}^{1/2}/(\rm{cm}^{1/2}\cdot \rm{s})$, this can be expressed 
as follows with $B$ in $\rm{G}$ (Gauss):
\begin{equation}
B < 9.158\cdot10^{-4} \cdot \left(\frac{R}{\rm{cm}}\right)^{1/2} \cdot \left(\frac{\sin\omega}{\rm{1}}\right)^{1/2} \cdot
    \left(\frac{|\mathbf{g}_{\rm{loc}}|}{\rm{g}/(s^2\cdot cm^2)}\right)^{1/2},                  \label{eq_collapse_criterion_numerical}
\end{equation}
where the numerical factor is absorbing the conversion of the gravitational constant to cgs units together with the factor
$\sqrt{4\pi}$, and $|\mathbf{g}_{\rm{loc}}|$ also needs to be expressed in cgs units as a gravitational force per volume.
Equation (\ref{eq_collapse_criterion_numerical}) can also be written as
\begin{equation}
B < 0.1913 \cdot \left(\frac{R}{\rm{mpc}}\right)^{1/2} \cdot \left(\frac{\sin\omega}{\rm{1}}\right)^{1/2} \cdot
    \left(\frac{|\mathbf{g}_{\rm{loc}}|}{\rm{M}_{\odot}\cdot\rm{M}_{\odot}/mpc^2}\right)^{1/2},                  \label{eq_collapse_criterion_numerical_2}
\end{equation}
where $B$ is in $\rm{G}$ (Gauss), and the numerical factor is again absorbing the gravitational constant, the factor
$\sqrt{4\pi}$, and the conversion to $\rm{M}_{\odot}$ and $\rm{mpc}$. $|\mathbf{g}_{\rm{loc}}|$ per volume is expressed in 
$\rm{M}_{\odot}\cdot \rm{M}_{\odot}/mpc^2$. For the final source-specific values we use a flux-to-total mass conversion of 2.15~mJy yielding 140~$\rm{M}_{\odot}$
at 1.3~mm with a dust temperature $\rm{T}_d=$100~K and a distance of 5.1 kpc.

%Folding in numerical values and adopting $\rm{cgs}$ units, this can be expressed as follows where $B$ is 
%expressed in $\rm{G}$ (Gauss):
%\begin{equation}
%B < N \cdot \left(\frac{R}{\rm{mpc}}\right)^{1/2} \cdot \left(\frac{\sin\omega}{\rm{1}}\right)^{1/2} \cdot
%    \left(\frac{|\mathbf{g}_{\rm{loc}}|}{\rm{M}_{\odot}/(s^2\cdot mpc^2)}\right)^{1/2}.                   %\label{eq_collapse_criterion_numerical}
%\end{equation}
%$N=2.3248\cdot 10^4$ is a numerical factor absorbing the conversion from the above units to the $\rm{cgs}$ units
%where $1\rm{G} = 1\rm{g}^{1/2}/(\rm{cm}^{1/2}\cdot \rm{s})$ together with the factor $(4\pi)^{1/2}$ and the gravitational constant used to derive
%$|\mathbf{g}_{\rm{loc}}|$. Using direct observables, equation (\ref{eq_collapse_criterion_numerical}) can be expressed as
%\begin{equation}
%B < N \cdot C \cdot \left(\frac{R}{\rm{arcsec}}\right)^{1/2} \cdot \left(\frac{\sin\omega}{\rm{1}}\right)^{1/2} \cdot
%    \left(\frac{|\mathbf{g}_{\rm{loc}}|}{\rm{(Jy/beam)}^2/(s^2\cdot arcsec^2)}\right)^{1/2},                 \label{eq_collapse_criterion_numerical_2}
%\end{equation}
%where $C$ is an additional numerical factor taking into account flux-to-mass conversion and angular scale ($\rm{arcsec}$)
%to physical scale at the source distance.  
%The here adopted dust continuum flux-to-total mass conversion is 2.15~mJy yielding 140~$\rm{M}_{\odot}$
%at 1.3~mm with a dust temperature $\rm{T}_d=$100~K and a distance of 5.1 kpc.

We note that the criterion in Equation (\ref{eq_collapse_criterion}) is fundamentally different from a mass-to-flux
ratio. The latter one uses a single field strength value to derive a magnetic flux over an area (volume) under 
consideration. Typically, this is applied to an entire core or cloud, and the magnetic field geometry is not taken 
into account. Equation (\ref{eq_collapse_criterion}) is making use of the resolved local magnetic field geometry, 
which can make a decisive difference as illustrated above with Figure \ref{figure_schematic_isolated_vs_filament}.
It is the direct comparison of the local direction of gravity versus the local direction of the magnetic field tension
force that leads to the criterion in Equation (\ref{eq_collapse_criterion}) while the mass-to-flux ratio is derived from 
an average magnetic field strength and an integrated mass.

Equation (\ref{eq_collapse_criterion}) defines a {\it maximum field strength that can be overcome by an observed 
field-gravity constellation} (the angle $\omega$ is measured and the magnitude of $|\mathbf{g}_{\rm{loc}}|$ can 
be derived with a dust-to-mass conversion factor), or it defines a {\it minimum field strength that can stabilize 
a dust lane or extension against a radial collapse}. 
Figure \ref{figure_stability_criterion} gives a breakdown of equation (\ref{eq_collapse_criterion}) by separately 
displaying maps for $\sin\omega$ and $|\mathbf{g}_{\rm{loc}}|$.
Field strength limits are of the order of a few tens of $\mu \rm{G}$ in the filamentary dust lanes and extensions, and they grow
to about 100-200~$\mu \rm{G}$ in the denser central regions. 
These field strength limits are smaller than any of the derived values based on dust polarization observations in 
W51 e2 and e8
\citep{tang09b,koch12a} or from OH maser measurements \citep{etoka12}
which are all of the order of a few $\rm{mG}$ up to about 10~$\rm{mG}$.
{\it This implies that these resolved filamentary dust lanes and extensions are unlikely to be able to collapse, and therefore, 
can form accretion channels stabilized by the magnetic field.}
This is supported by recent results from \citet{goddi20} that also find filamentary dust lanes
in W51 e2-E and e8 (at 1.3~mm with a resolution of 0\farcs02, but without magnetic field detection) which are interpreted as accretion flows.

%B-field mostly aligned with fibers in plane of sky also means that with high probability 
%B-field is aligned in 3D (BISTRO paper, Tomisaka)? 

\subsection{Applicability}
\label{section_applicability}

In this section we address possible complications in the method and criterion introduced in the above sections \ref{analysis_local_gravity} and \ref{analysis_B_induced_stability}. 
While illustrated specifically with observed quantities for 
W51 e2 and e8, the following considerations and line of argument remain generally valid. 

{\it Opacity.} Dust around 230~GHz ($\sim1.3$~mm) can become optically thick in the 
very centers of high-mass
star-forming regions. Our starting point in section \ref{analysis_local_gravity} was to take an observed
dust distribution as a tracer for the distribution of the overall mass. If dust is indeed optically thick 
in the very center region, this assumption will not hold for that very region as some mass (dust and gas) might be undetected. To be specific, for W51 e2, 
the recent work by \citet{goddi20} finds that dust continuum (observed with ALMA around 1.3mm with 
a resolution of about 0\farcs02) is optically thick up to a radius $\lesssim$ 1000 au in W51 e2-E with a 
brightness temperature $\gtrsim$~200~K.
This is in agreement with simulations that predict that dust can become optically thick for such central regions \citep{forgan16,klassen16}.
Depending on opacity, \citet{ginsburg17} find a mass of 18 $M_{\odot}$ per beam for the central region in e2-E in the optically thick case ($\tau\ge1$) and about 6 $M_{\odot}$ per beam in the optically thin case ($\tau\lesssim 1/3$) where one beam refers to their 0\farcs2 ALMA observations around 1.3mm.
The later estimates in \citet{goddi20} with a beam of 0\farcs02 are 9.7 $M_{\odot}$ for the central core object in e2-E within a radius of about 500 au, if entirely optically thick, and about 4 $M_{\odot}$ if optically thin. Larger estimates of about 12 $M_{\odot}$ result for lower dust temperatures. 
Hence, for these two different resolutions, the mass estimates for the central region differ roughly by a factor 2 to 3. These estimates are based on uniform filling factors, and unknown substructures within 
the beam cannot be accounted for. A similar range in mass estimate is found for the central region covered by one 0\farcs2 beam in W51 e8 \citep{ginsburg17}. 
{\it How does this uncertainty in central mass, resulting from these opacity effects, impact our calculation of the gravitational vector field in Figure \ref{figure_loc_grav}?}
The effect is likely very minimal for the following reasons.
First, the optically thick central areas with a radius 500 to 1000 au (0\farcs1 to 0\farcs2) are very small compared to the full extension and area ($\sim3\arcsec \times 3\arcsec$) for both W51 e2 and e8. These central areas are only covering 4 to 16 beams in our observations with a resolution of 0\farcs1.
Second,  the gravitational vector field is calculated taking into account the full area. Since the bulk of 
the mass is still in the extended area, a factor of 2 to 3 difference in a relatively small mass concentrated
in the very center is negligible. In other words, the gravitational effect of this difference is largely diluted
over the entire W51 e2 and e8 areas. 
Third, for both e2 and e8, the gravitational vector field towards and inside the central 0\farcs2 area
is very radial (Figure \ref{figure_loc_grav}). A larger or smaller mass concentration (within a factor of a few) will not change the {\it direction} of local gravity, because the dust distribution is becoming increasingly circular in the very center, and hence the alignment measure $\sin\omega$ is unaffected. The {\it magnitude} of local gravity
might change, in a most conservative estimate by a factor of 2 to 3,  but more likely by significantly less because it will again be diluted by the distribution of the extended area outside of the very central region.

{\it Outflows.} The bottom right panels in Figure \ref{figure_B} and \ref{figure_B_2} overlay
 the small-scale collimated SiO 
outflows \citep{goddi20} and the 
wider larger-scale $^{12}$CO outflows \citep{ginsburg17} on our dust continuum and magnetic field detection for W51 e2-E and e8. For e2-E, both outflows are clearly bipolar, 
they are both along a main southeast-northwest direction, 
%the SiO outflow is possibly driving the larger-scale $^{12}$CO outflow, 
and possibly both outflows have the same origin. 
W51 e8-N displays a small-scale SiO outflow along an east-west direction and a larger-scale $^{12}$CO
that appears less regular.
{\it How far is the presence of these outflows affecting our analysis?}
Overall, the morphologies in dust and outflows and their spatial locations appear rather different and 
uncorrelated for both W51 e2 and e8, i.e., 
the SiO outflow appears only within a central area of about 0\farcs4 or less,
covering only a small fraction of 
the area over which the B-field is mapped. Furthermore, the B-field morphology across both the red- and blue-shifted SiO lobe is regular and coherent without any feature that would distinctively delineate
a boundary of the SiO outflow. Similarly, no distinct feature in polarization appears at these locations
(Figure \ref{figure_polarization}). The mass estimate for the e2-E outflow is $0.36~M_{\odot}$ \citep{goddi20}, which is negligibly small 
compared to the central mass and all the extended emission. With a mass of $0.11~M_{\odot}$, the 
same holds for the SiO outflow in e8-N.
The $^{12}$CO outflows are significantly more extended, 
with the northern lobe in W51 e2 reaching out to e2-NW
and the southern lobe extending south far beyond our detected B-field structure, 
and with both red- and blue-shifted lobes covering most of the mapping area in W51 e8.
Analogous to the SiO outflow, the B-field does not show any delineating features between the inside- and outside-outflow areas, but appears coherent and connected across the entire W51 e2 and e8 systems.
We note that the brightest contours in the southern $^{12}$CO lobe in e2-E might fall in between dust lane 1 and 2 (Figure \ref{figure_B}, bottom left panel). Similarly, the southern SiO outflow is pointing
at the void between dust lane 2 and 3, and the rim of this void shows relatively high
polarization fractions (Figure \ref{figure_polarization}).
This could 
indicate that some of the dust lanes are being additionally shaped by the outflows, besides gravity. For a boundary layer
between dust lane and outflow, local gravity could then not be calculated as we propose, because the outflowing material will entrain the dust in this layer.  However, for the bulk of material inside the dust lane, away from the boundary layer, local gravity can still be calculated as suggested. 
Moreover, the majority of the dust lanes do not seem to be impacted by the outflows. This finding 
is consistent with a conclusion in \citet{goddi20} where they argue that the dust lanes on a scale 
of a few thousand au (with a resolution of 0\farcs02) are accretion flows and not outflows. 
Adding that overall both W51 e2 and e8 are gravity-dominated, and generally very little dust is 
expected to be present in outflow cavities (i.e., dust is likely too weak to be seen), 
we conclude that the method and criterion in 
sections \ref{analysis_local_gravity} and \ref{analysis_B_induced_stability}
are applicable in the presence of these outflows.
As the magnetic field is stabilizing the dust lanes against collapse, it is at the same time also stabilizing them 
against any external pressure.

{\it Radiative Feedback.}
Both thermal and ionizing radiation are expected in high-mass systems. For W51 e2 and e8, 
\citet{ginsburg17} indeed identify chemically enhanced regions as a result of radiative feedback 
heating the molecular gas. These regions are likely heated by direct infrared radiation from newly 
forming stars in their centers. With a resolution of $\sim0\farcs2$, these heated regions display
morphologies that are overall similar to their detected continuum emission, although regions in different
molecular lines vary in their sizes from more compact to more extended. This suggests that 
in these sources on this scale, the dominating effect of this radiative feedback is mostly symmetrical heating and not any disruptive event which would lead to more irregular, disconnected, and broken up morphologies both in these lines and the magnetic field. 
We note that, e.g., in the presence of expanding HII regions with swept-up shells, the dynamical constellation could be very different. This is observed in the high-mass system G5.89--0.39 which precisely 
shows swept-up shell morphologies in continuum with a magnetic field aligned with its expanding front
\citep{tang09a,fernandez21}.
We do not find any indication of such features in W51 e2-E or e8. There might be a hint of a partial 
shell-like B-field morphology around e2-W, which indeed is harboring an HII region. In this case, dust 
lane 6 (Figure \ref{figure_B}, bottom left panel) 
might indeed be affected by this feedback, though the polarization coverage is 
incomplete to be fully conclusive.
Hence, we conclude that unless imprints from feedback are clearly visible in continuum and B-field 
morphology, these systems are still gravity-dominated and our approach remains valid. Radiative 
heating might further stabilize accreting dust lanes against fragmentation and local collapse.
In the case of a dynamically overwhelming feedback, the collapse criterion would need to be expanded
with an additional pressure gradient term that can be added to the local gravitational force.

\section{Discussion: Towards a Comprehensive Picture from Large-Scale Filamentary Envelope to Core-Accreting Dust Lanes} 
\label{section_discussion}

Combining with earlier data sets, covering physical lengths from about 0.5~pc to 2.6~mpc with resolutions $\theta\sim3\arcsec$ to $0\farcs1$ 
(i.e., going through a range of almost 1,000 in resolved area), 
we are able to isolate B-field structures in 4 distinct scales and regimes 
where the B-field is playing different roles. 
The following sections summarize these different regimes (Figure \ref{figure_synergy_1}) and diagnostic tools 
(Figure \ref{figure_synergy_2}), and discuss their implications. 
Figure \ref{figure_visualization} visualizes the evolving role of the B-field across these regimes.

\subsection{Distinct Scales: Envelope -- Global Collapse -- Local Collapse -- Core-Connecting Dust Lanes}
\label{discussion_synergetic_picture}

%based on tailored analysis approaches to access the role of the magnetic field. 

\subsubsection{Envelope-to-Core Scale.}   \label{discussion_envelope_to_core}
On this largest scale, the W51 e2 and e8 cores are together embedded in a $\sim 0.5$~pc elongated north-south envelope 
(Figure \ref{figure_synergy_1}, left and second left panel).
The histogram of magnetic field P.A.s \citep[SMA subcompact, $\theta\sim2\arcsec$;][]{koch18}
indicates a prevailing field orientation perpendicular to the envelope's longer axis. 
The e2 and e8 cores are aligned perpendicular to this dominating east-west B-field orientation. 
This B-field-envelope configuration is suggestive of accretion from the east and from the west.\footnote{
We note that the W51 North complex is displaying analogous features on this envelope scale, revealing B-field structures suggestive
of accretion from north and south. All the smaller cores within W51 North align along an east-west direction, 
perpendicular to the suggested accretion direction \citep{koch18, tang13}. 
}
Despite a prevailing field orientation, first twists are visible in the B-field morphology (Figure \ref{figure_synergy_1}, left panel). 
An analysis based on a local magnetic field dispersion map\footnote{
The local B-field dispersion captures by how much a local field orientation varies with respect to its surrounding.
Generally, maps can be generated with respect to fewer or more surrounding pixels (beams) capturing a smaller or 
larger scale over which the B-field orientation varies. Maps will display larger dispersion values where 
the B-field bends more rapidly or changes orientation abruptly \citep{koch18, fissel16, planckXIX, planckXX}.
} 
shows wide areas of very small dispersion indicating
and confirming a prevailing B-field orientation, but also a few areas of significant local dispersion. These larger-dispersion
areas occur along a southeast-northwest axis in e2 and at the southern end of e8
(Figure \ref{figure_synergy_2}, left panel). 
The predictive measure of this analysis on this scale is that {\it these specific locations are pinpointing to 
where and along which direction the global collapse in e2 and e8 is happening.} This defines location and scale
of the initial gravitational drag towards the forming e2 and e8 cores. 
Zooming out, the BIMA observations (\citet{lai01}; $\theta\sim3\arcsec$, left panel in 
Figure \ref{figure_synergy_1}) possibly probing the outer surface
of this envelope, reveal almost uniform B-field orientations
along an east-west direction except for the e2 northwestern corner that shows a hint of bent field lines towards e2. 

%Displaying analogous features, the W51 North complex on this envelope scale reveals B-field structures suggestive
%of accretion from north and south. All the smaller cores within W51 North align along an east-west direction, 
%perpendicular to the suggested accretion direction \citep{koch18, tang13}. 

A second important feature on this envelope-to-core scale is the detected bridge between e2 and e8 
(around Dec. offset $-3$ in the second panel in Figure \ref{figure_synergy_1}). 
While the southern tip of e8 shows a first glimpse of a gravitationally dragged-in 
B-field morphology, its northern end reveals field lines that are gradually deflecting towards the more massive e2.
This clearly different morphology from one end to the other end of the core -- as opposed to the symmetrical 
morphology in the more massive e2 core -- is interpreted as local collapse starting at the southern
end of e8 while the northern end cannot locally collapse but is being pulled to the more massive e2 core
(also see later discussion in section \ref{discussion_multi_scale_collapse} and Figure \ref{figure_self_similar_structures}). 
This particular feature is analyzed in a forthcoming work that investigates gravitational entrainment
of the B-field in order to derive a field strength estimate (Koch et al., in preparation).

\subsubsection{Global-Collapsing-Core Scale.}  \label{discussion_global_collapsing_core}
On this $\sim0.05$pc scale the e2 and e8 cores are clearly detected as entities, the surrounding larger-scale diffuse envelope is resolved out, 
and in particular e2 shows an hourglass-type B-field morphology towards its dominating gravitational center. The 
field lines are clearly bent and almost radial-like \citep[SMA extended, $\theta\sim0\farcs7$; ][]{tang09b}.
The B-field detection in e8 is less complete though also hinting collapse with field orientations directed towards
the main center in e8 (Figure \ref{figure_synergy_1}, middle panels).
With the introduction of the angle $\delta$ as an observable and diagnostic 
tool\footnote{
The measurable angle $\delta$ between a local magnetic field orientation and an intensity gradient is 
introduced in \citet{koch12a}. Its merit as a key diagnostic is discussed in \citet{koch13}. 
In particular, $\delta$ is an approximation to $\Sigma_{\rm{B}}$, because changes in $\delta$
closely correlate with changes in $\Sigma_{\rm{B}}$. Additionally, $\delta$ can be given a sense of 
orientation, i.e., magnetic field orientations can be rotated clockwise or counter-clockwise with respect to
an intensity gradient. This, e.g., leads to a characteristic bimodal distribution in $\delta$ for hourglass-type
B-field morphologies, such as for W51 e2. The regions transitioning between $+\delta$ and $-\delta$
are signposts for the zones of accretion and outflow \citep{koch13}.
}
, the polarization--intensity gradient method leads to a $\Sigma_{\rm{B}}$-map for e2 with values around 0.2 to 
0.3 across the entire core. $\Sigma_{\rm{B}}$ values below 1, i.e., a magnetic field tension-to-gravity force ratio
below 1, indicate that gravity is overwhelming the B-field in this area. {\it The e2 core, as an entity, can thus engage in 
a global collapse} (Figure \ref{figure_synergy_2}, second left panels).
It is only the northwestern extension where $\Sigma_{\rm{B}}$ is larger than 1 or around 1. Here, the B-field
is strong enough to dominate over gravity and halt any infall or collapse motion. Looking at the radial profile 
of $\Sigma_{\rm{B}}$ there is an obvious drop from values around 0.8 at the periphery of the core to values
of 0.2 in the center of the core. Consequently, {\it the magnetic field in the observed configuration is leading to a gravity
dilution on the global-collapsing-core scale,
i.e., a gravity efficiency smaller than 1, quantified by $\Sigma_{\rm{B}}$}. Gravity is only about 20\%
effective in the core's periphery while quickly increasing to around 80\% in the inner core as compared 
to free-fall enabled by gravity exclusively \citep[Figure 4 in][]{koch12b}.
In an analogous way, the profile of mass-to-flux ratio, independently derived from $\Sigma_{\rm{B}}$, 
shows a transition from subcritical at larger radius to supercritical within the core \citep[Figure 6 in][]{koch12b}.
The magnetic field strength map, derived with the polarization--intensity gradient method, displays a field strength
$B$ growing from the periphery to the center, reaching around 19~mG in the center with a radial profile close to 
$B(r)\sim r^{-1/2}$. 
It is evident from this analysis that the magnetic field properties are not constant across the core, but 
vary substantially in defining the role of the magnetic field (Figure \ref{figure_synergy_2}, second left panels).
It is interesting to note that the orientation of the e2 core with its northwestern extension is 
clearly reflected with the axis of larger B-field dispersion values (on the envelope-to-core scale) as noted above
in section \ref{discussion_envelope_to_core}.

We remark that identical trends, as seen in $\Sigma_{\rm{B}}$ and mass-to-flux ratio on this global-collapsing-core scale, are 
also seen on zoomed-out larger scales, namely a field-to-gravity force ratio $\Sigma_{\rm{B}}$ being largest {\it in between 
e2 and e8} with values dropping towards both cores, and again on an even further zoomed-out scale,  $\Sigma_{\rm{B}}$ 
being largest {\it in between 
W51 North and W51 e2/e8 together} with values dropping towards the W51 e2/e8 complex
and the W51 North complex \citep{koch12b}.
This suggests a self-similar collapse picture, or multi-scale collapse-within-collapse scenario as 
discussed further in section \ref{discussion_multi_scale_collapse}.

\subsubsection{Local-Collapsing-Core Scale.}    \label{discussion_local_collapsing_core}
Zooming in on e2 and e8 reveals detailed substructures below the above global-collapsing-core scale in section
\ref{discussion_global_collapsing_core}.
The cores fragment into smaller cores and with that generic B-field morphologies are detected \citep{koch18}.
The previous {\it hourglass-type global-collapse signature is resolved into smaller-scale features reflecting
local and inter-core dynamics imprinted onto the B-field morphology} (Figure \ref{figure_synergy_1}, 
second right panels).
In particular, 
satellite cores with a bow-shock shaped field structure are seen in e2-NW and e8-S. They both appear
to fall into or merge with their bigger neighboring cores, e2-E/e2-W and e8-N, respectively. 
From two sides centrally converging B-field structures are visible along a northwest direction 
from e2-E and in between e8-S and e8-N. 
Several cores reveal a {\it prominent B-field asymmetry} -- gravitationally bent and dragged-in B-field morphology in one
half of a core with compressed, straightened and deflected field lines in the other half of the core -- {\it suggestive of local 
collapse happening on one 
end while the core itself is pulled towards the next bigger more massive neighboring core.} 
This is seen in e2-NW being pulled towards e2-E/e2-W, and e8-S being pulled towards e8-N.\footnote{
We note that analogous features are also seen in W51 North, with N2 being pulled towards N1
and N4 being pulled towards N3 \citep[Figure 2 panel (h) in][]{koch18}.}

%{\it This local collapse happening simultaneously with the above described global collapse points at a
%collapse-within-collapse scenario.}
%smaller-scale locally collapsing cores within a larger-scale globally collapsing core -- is seen in LIST
%Recent numerical work also suggests a multi-scale, global hierarchical collapse scenario \citep{vazquez19}, 
%consistent with our multi-scale observational findings of B-field morphologies. 

This locally varying role of the B-field, specifically focusing on the inter-core and converging field structures,
is reflected in the $\sin\omega$ measure.  $\sin\omega$ in the range between 0 and 1 measures the fraction of 
the local magnetic field tension force that can work against local gravity. The asymmetrical B-field structure
in these locally collapsing cores is captured with $\sin\omega$ values around 0 on the collapsing sides 
and with maximum $\sin\omega$ values close to 1 on the sides being pulled to the next larger collapse
centers (Figure \ref{figure_synergy_2}, second right panels).
{\it Thus, on this local-collapse and inter-core scale the role of the B-field can flip from essentially no 
resistance to maximum resistance against collapse. }
Similarly, low $\sin\omega$ values across the global e2 and e8 cores localize regions where gravity 
is unobstructed.  These regions were suggested to represent magnetic channels in \citet{koch18}.

\subsubsection{Accreting-Dust-Lane Scale.} 
Moving to a $0\farcs1$ resolution ($\sim540$~au) seems to resolve a critical physical
length scale around the W51 e2 and e8 core in such that we start to witness the departure from near-spherical
structures and resolve a network of filamentary dust lanes and extensions
(Figure \ref{figure_B}, \ref{figure_B_2}, and right panels in Figure \ref{figure_synergy_1}).
As described in the 
above sections \ref{analysis_local_gravity} and \ref{analysis_B_induced_stability}, the observed 
characteristics on this scale are (1) a B-field prevailingly aligned with dust lanes, and (2) 
local gravity mostly directed along dust lanes towards the central mass concentration. 
Given this observed morphology, section \ref{analysis_B_induced_stability} is investigating the B-field-induced
stability against a radial collapse in these dust lanes. Since it is found that small field strength values 
are already sufficient to balance local gravity (these values are smaller than previously derived and observed values),
it is unlikely that these structures further fragment and collapse. 
This leads to the conclusion that the {\it magnetic field is playing a decisive role in stabilizing these structures}. 
Based on this, we propose that these observed dust lanes represent stable 
accretion zones and channels towards the 
denser core regions where material is guided by magnetic fields. 
Adding the magnetic field into this picture complements recent results by \citet{goddi20} who 
interpret the filamentary dust lanes arising from the compact cores in W51 e2-E, e8, and North
as multidirectional mass accretion.
If indeed present, these accretion streams can provide a mechanism to feed 
100-au-scale or smaller
Keplerian disks. 
Such a mechanism can be of interest because the formation of large and stable 1000-au-scale disks in high-mass star-forming systems can be challenging \citep[e.g.,][]{rosen19,seifried15,myers13, commercon11}.
Further support for such a scenario comes from the numerical work by \citet{mignon21} who find filamentary 
structures linking to magnetically regulated $\sim$100 to 200-au disks. These structures, referred to as streamers, 
are identified as a path for accretion flows where the magnetic field is pulled along, and the gas moves along 
the field lines. Their accretion streamers are thermally dominated and magnetically stabilized.

\begin{figure}
\includegraphics[width=9cm]{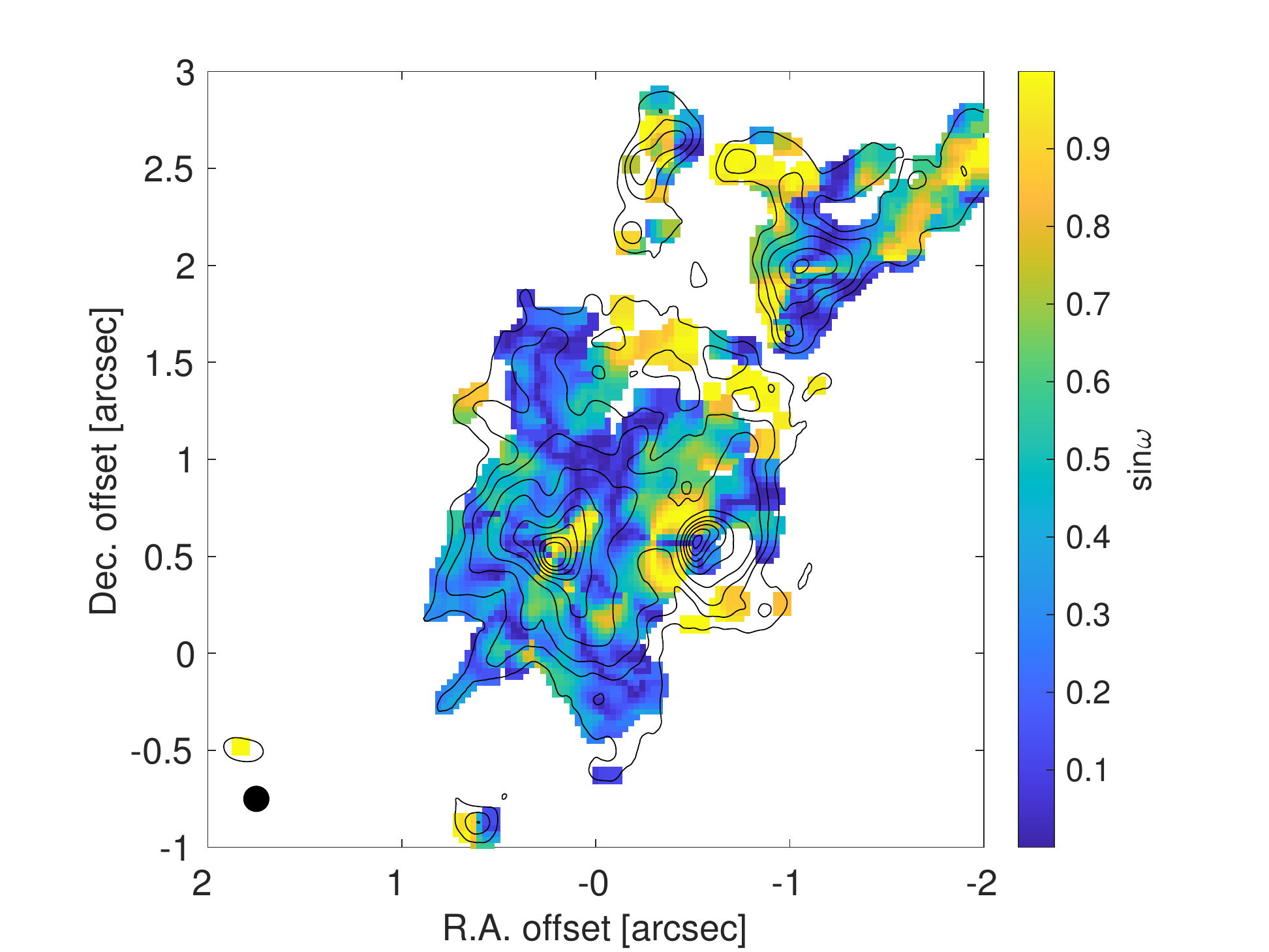}
\includegraphics[width=9cm]{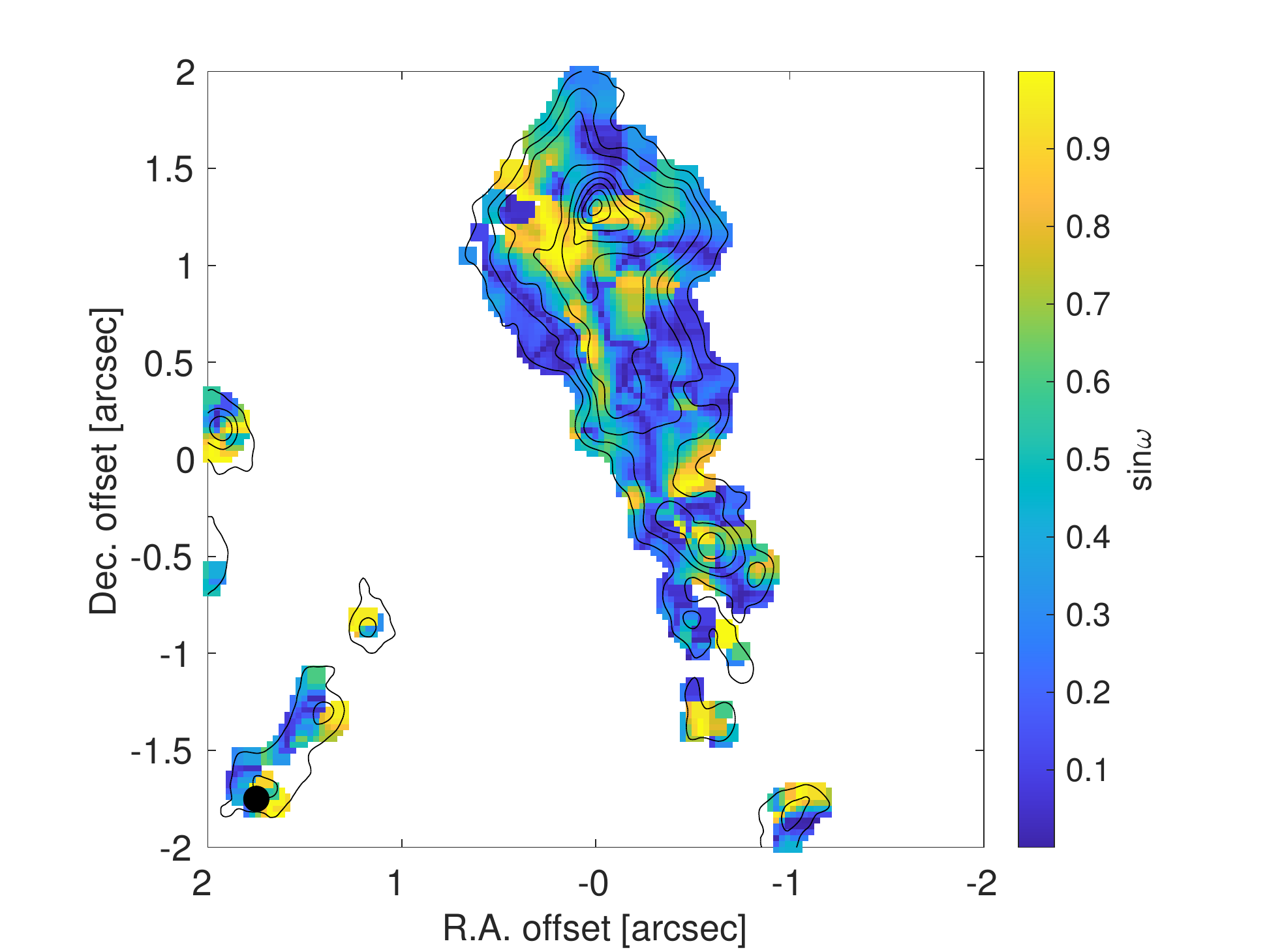}
\includegraphics[width=9cm]{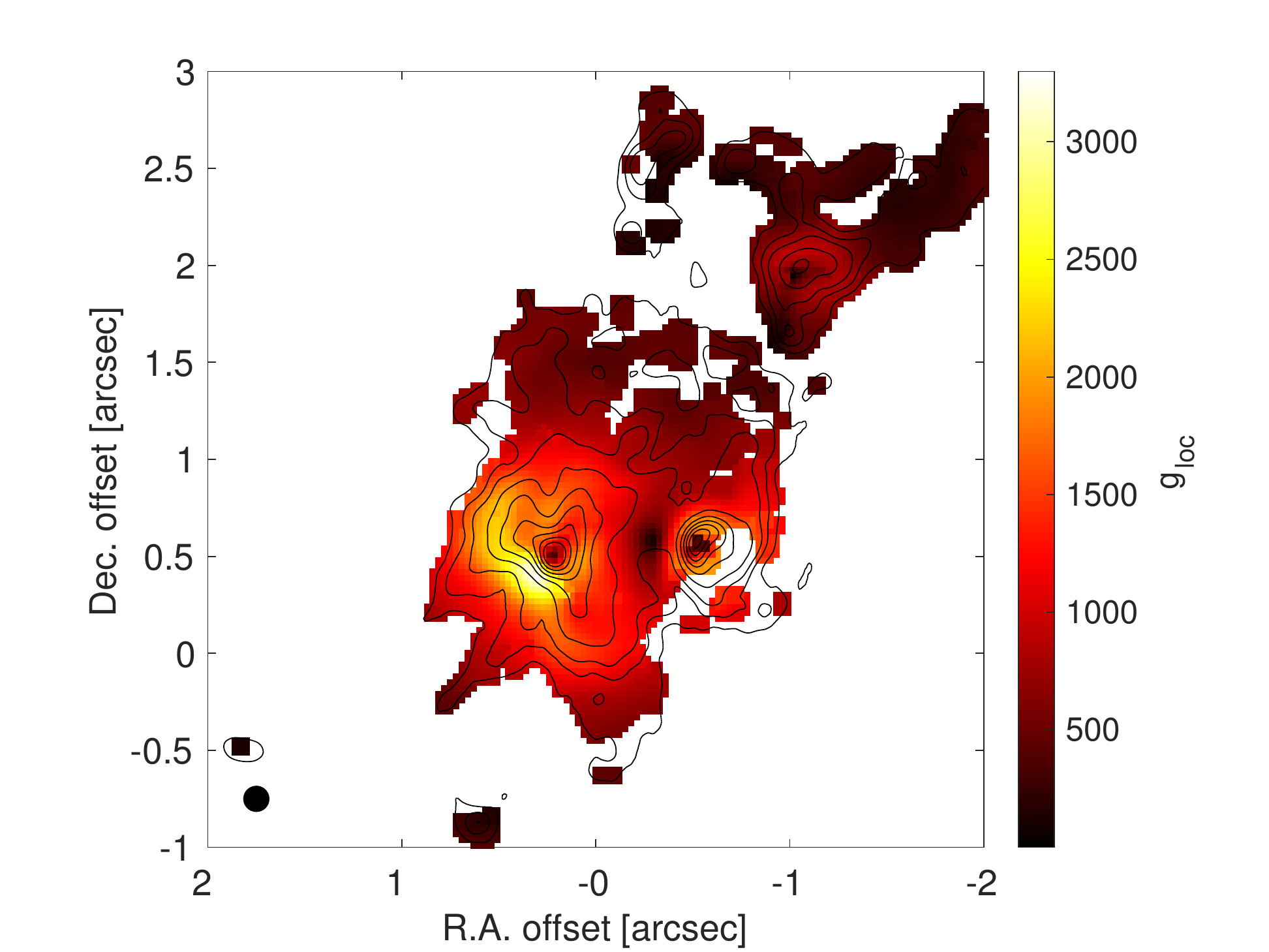}
\includegraphics[width=9cm]{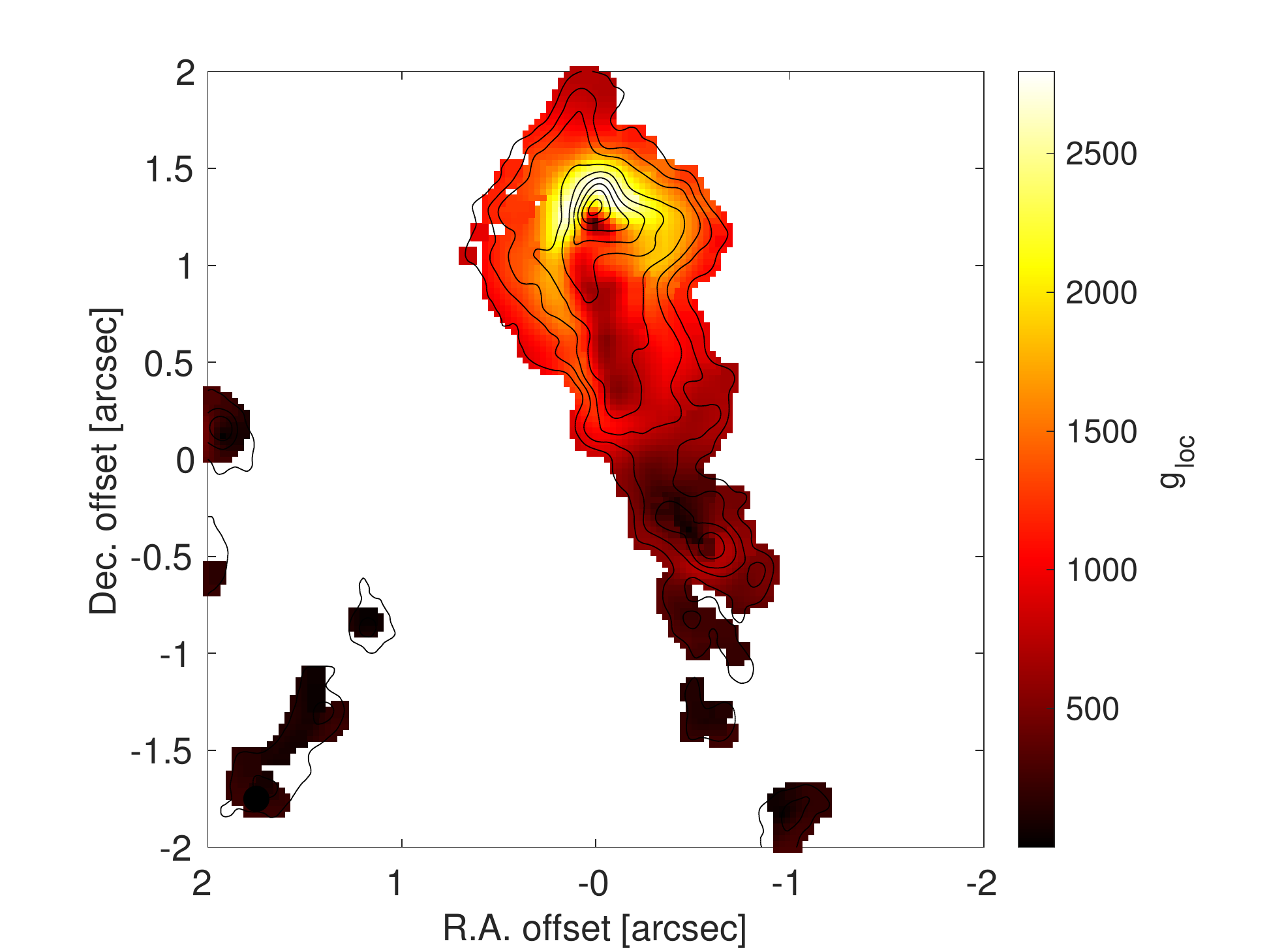}
\includegraphics[width=9cm]{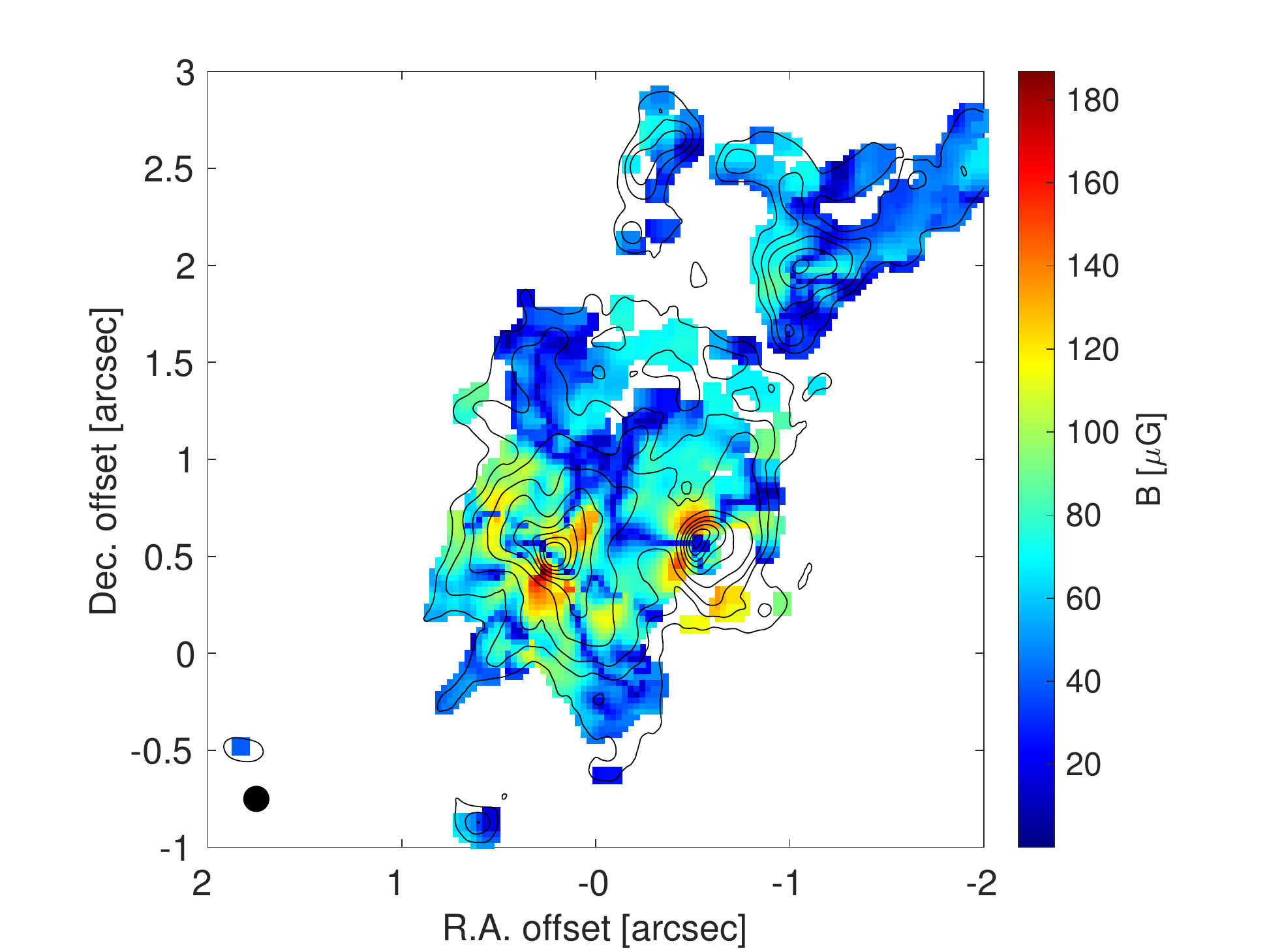}
\includegraphics[width=9cm]{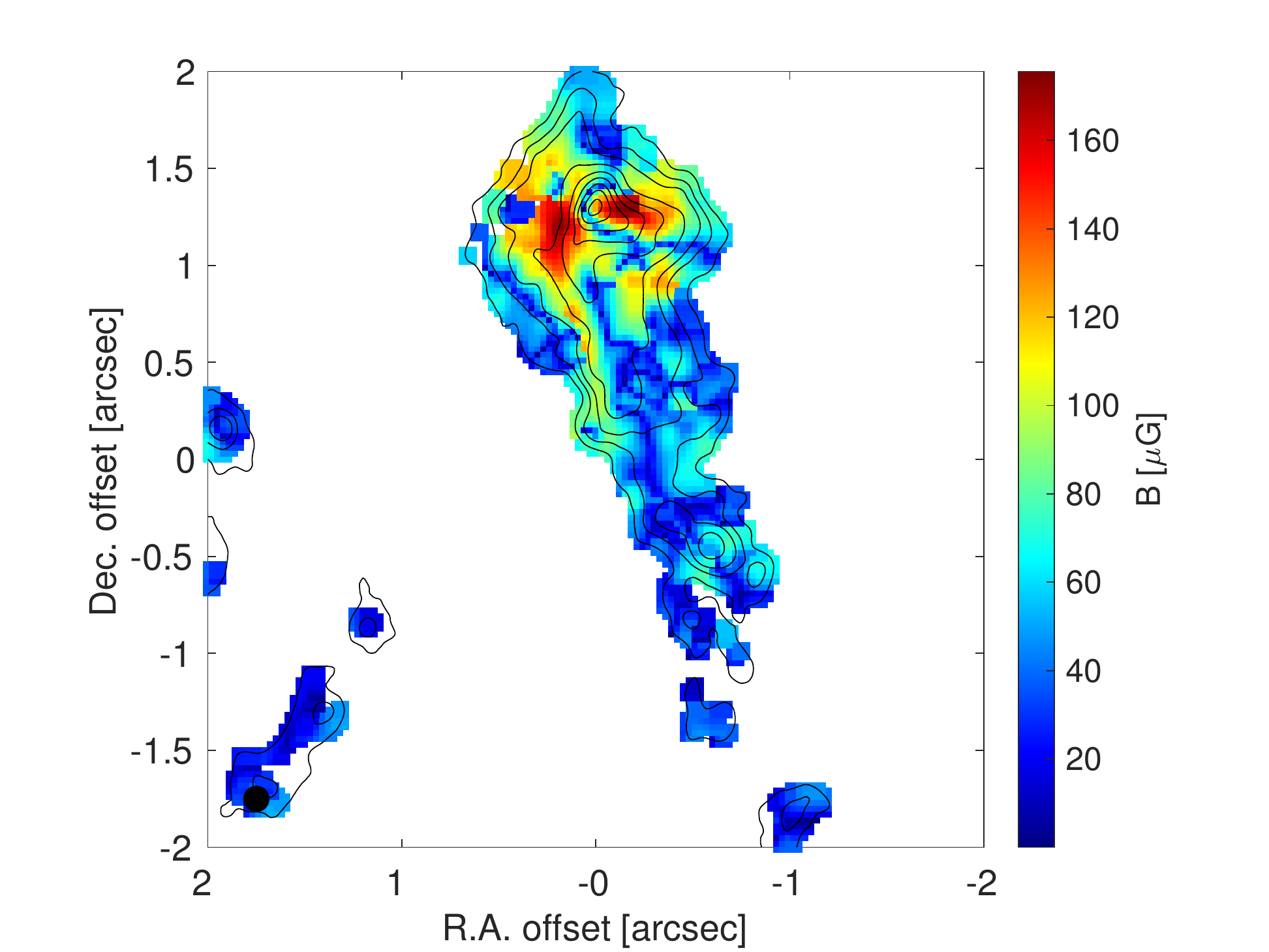}
\caption{Local magnetic field strength maps (bottom row) derived from the stability and collapse criterion given in equation (\ref{eq_collapse_criterion}).  
Separately displayed are the measurable input parameters $\sin\omega$ (top row) and $|\mathbf{g}_{\rm{loc}}|$ (middle row; arbitrary units) 
for W51 e2 (left column) and e8 (right column).
}
\label{figure_stability_criterion} 
\end{figure}

\begin{sidewaysfigure}
\includegraphics[trim=0cm 9cm 0cm 0cm, clip=true,width=22cm]{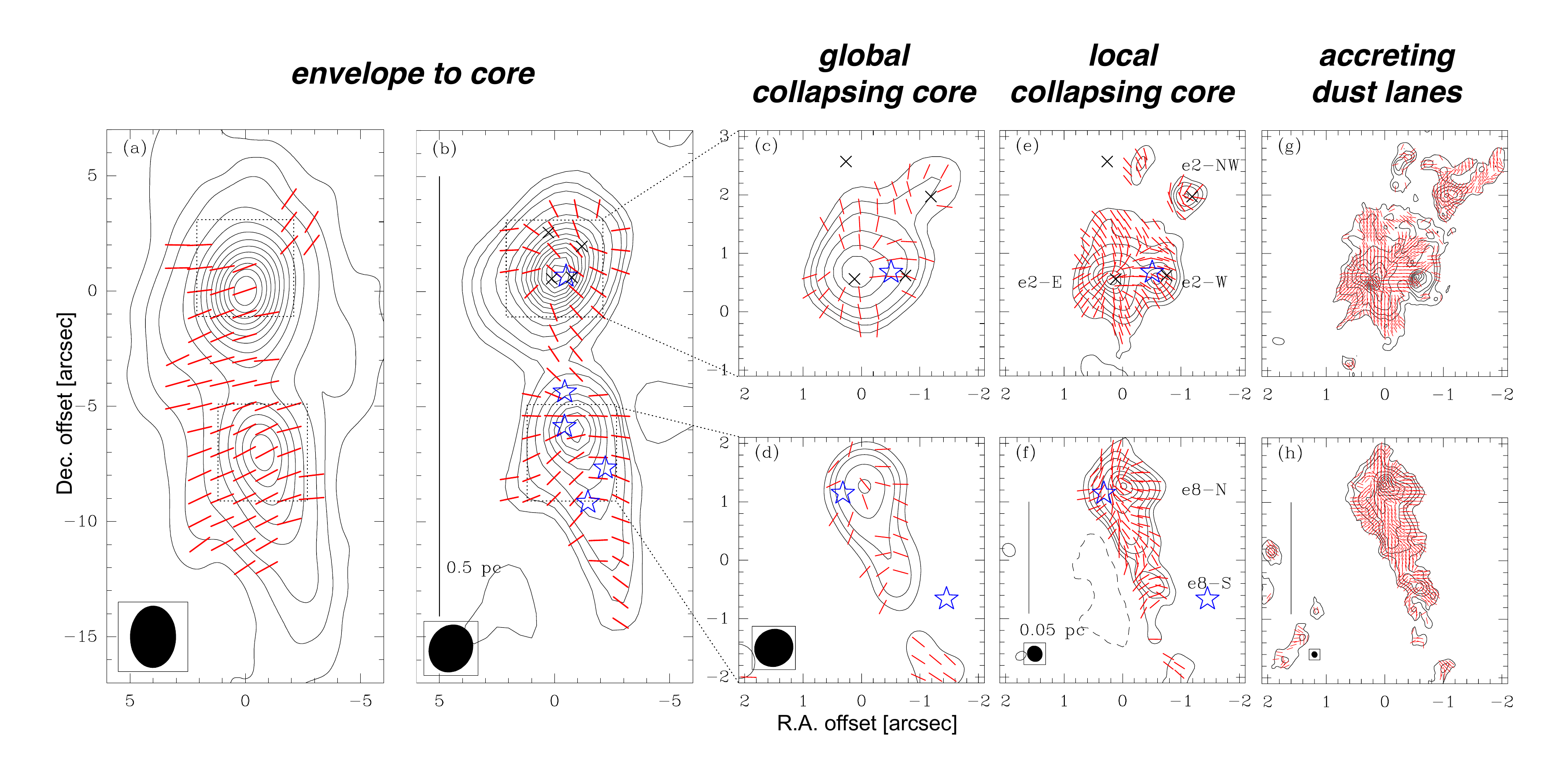}
\caption{\small Overview of B-field structures (red segments) in the W51 e2/e8 region with resolutions from about 
$\theta\sim3\arcsec$ to $0\farcs1$, 
covering scales from about 0.5~pc to 540~au. Contours are dust continuum. Panels from left to right display the distinct regimes as 
discussed in section \ref{discussion_synergetic_picture}: envelope-to-core scale, global-collapsing-core scale, local-collapsing-core scale, 
and accreting-dust-lane scale.
In panel (a) contours are 3, 6, 10, 20, 30, 40,...$\times\sigma$, where $1\sigma$ is 27 mJy/beam 
with $\theta\sim2\farcs7\times2\farcs0$ 
at a wavelength $\lambda$ around 1.3~mm.
Contours are 3, 6, 10, 20, 35, 50, 65, 80, 95,...$\times\sigma$, where $1\sigma$ is 75 mJy/beam 
with $\theta\sim2\farcs13\times1\farcs88$ 
and $\lambda\sim0.87$ mm
in panel (b), 
60 mJy/beam with $\theta\sim0\farcs7\times0\farcs6$ 
and $\lambda\sim0.87$ mm
in panel (c) and (d), 
6 mJy/beam with $\theta\sim0\farcs28\times0\farcs26$ 
and $\lambda\sim1.3$ mm
in panel (e) and (f), 
and 1.2 mJy/beam 
with $\theta\sim0\farcs11\times0\farcs10$ 
and $\lambda\sim1.3$ mm
in panel (g) and (h).
Panels (b) to (f) are adapted from \citet{koch18}.
}
\label{figure_synergy_1} 
%\end{figure}
\end{sidewaysfigure}

\begin{sidewaysfigure}
\includegraphics[trim=0cm 7cm 0cm 0cm, clip=true,width=22cm]{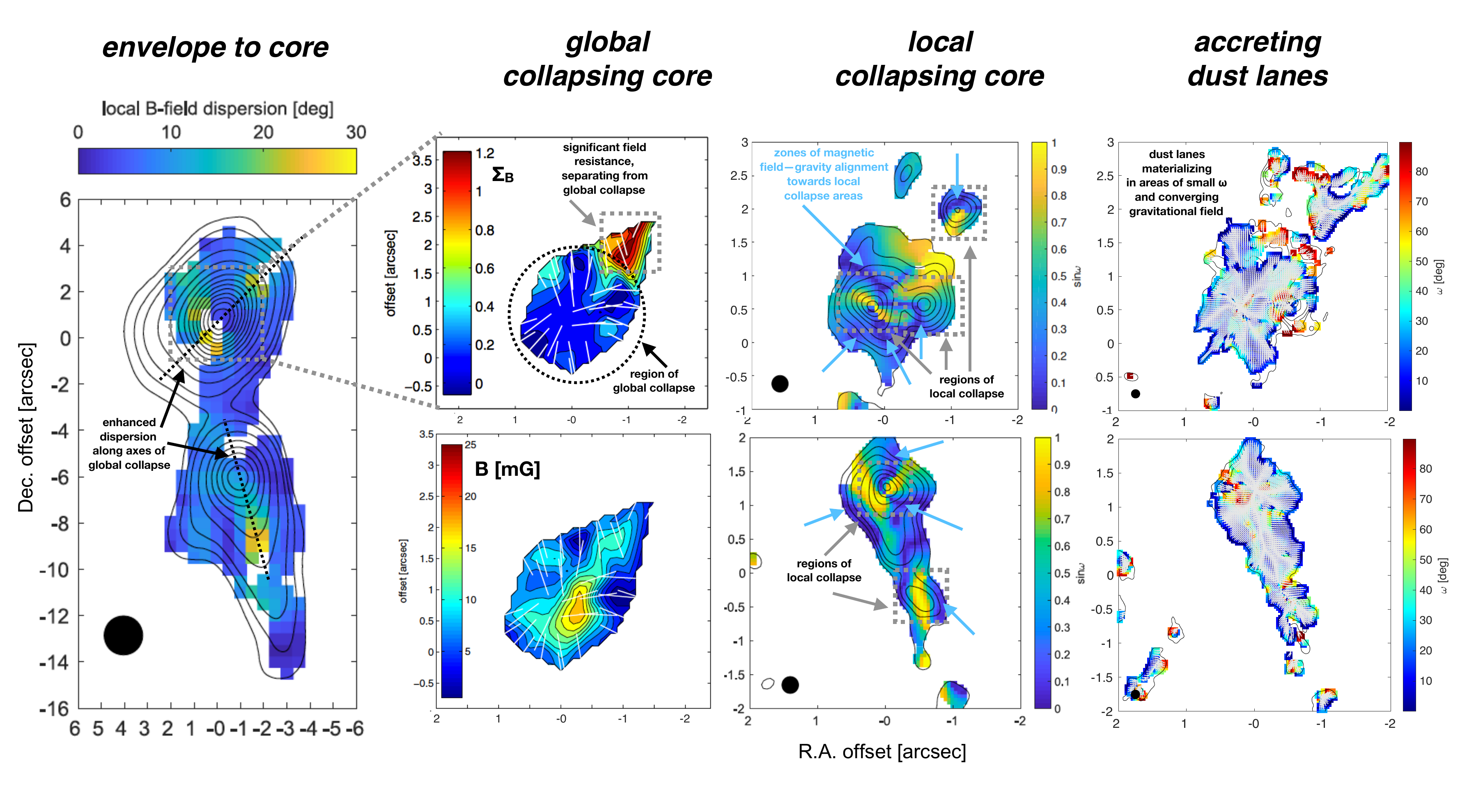}
\caption{\footnotesize Overview of diagnostic tools in the W51 e2/e8 region. 
From left to right, on the envelope-to-core scale the local magnetic field dispersion can pick out location and direction
where a global collapse will start to happen (zones with larger dispersion values); 
on the global-collapsing-core scale, the small values of the magnetic field tension-to-gravity force ratio $\Sigma_{\rm{B}}$ 
across the core indicate that e2 as an entity can globally collapse (upper panel), and the measured magnetic 
field strength map reveals a clearly growing strength towards the center (lower panel); 
on the local-collapsing-core scale, the $\sin\omega$ measure locates regions where gravity is unobstructed (small values)
and where the B-field can maximally slow down gravity (values close to one) towards locally collapsing smaller cores;
on the accreting-dust-lane scale (rightmost panels), ridges formed by the locally converging gravitational field (gray arrows) appear in areas labelled as dust lanes in Figure \ref{figure_B} and \ref{figure_B_2}, which typically also
coincide with small angles $\omega$ (color scale).
Low $\sin\omega$ values on the local-collapsing-core scale are signposts of the appearance of dust lanes at higher resolutions.
Panels in the first three columns are adapted from \citet{koch12a} and \citet{koch18}.
}
\label{figure_synergy_2} 
%\end{figure}
\end{sidewaysfigure}

\begin{sidewaysfigure}
\includegraphics[trim=0.5cm 0.5cm 1cm 12.2cm, clip=true,width=22cm]{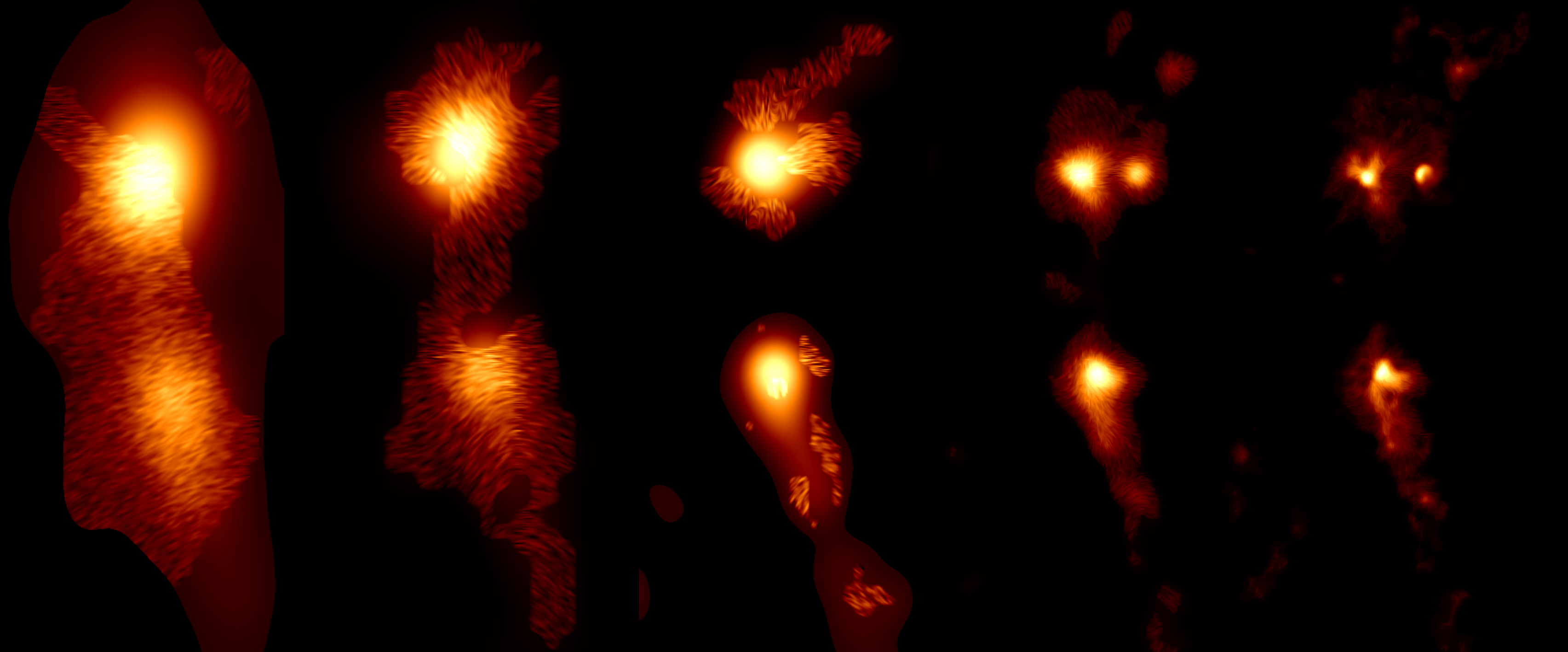}
\caption{\small Visualization of the evolving roles of the magnetic field over the scales presented in Figures 
\ref{figure_synergy_1} and \ref{figure_synergy_2}.
Visualized are streamlines based on the detected magnetic field morphologies in Figure \ref{figure_synergy_1}
using a line
convolution algorithm for vector fields (e.g., \citet{cabral93}).
From left to right are envelope-to-core scale (first and second), 
global-collapsing-core scale (third),
local-collapsing-core scale (forth), and accreting-dust-lane scale (fifth).
No visualization is shown in black patches with too sparse sampling in the original data
which does not allow to construct a connected convolution. 
}
\label{figure_visualization} 
\end{sidewaysfigure}

Recent work on the Serpens South region has also found a parallel B-field alignment in one filament that is interpreted as
evidence for gravity dragging the denser gas, in this case derived in the context of a change of alignment as compared to lower 
visual extinction areas,  and entraining the flux-frozen large-scale B-field \citep{pillai20}.
Calculating the actual gravitational vector field (Figure \ref{figure_loc_grav}) adds an important observable and 
strongly corroborates this picture in W51 e2 and e8.
The $\sin\omega$ maps (top row in Figure \ref{figure_stability_criterion}) 
further substantiate this, clearly displaying very small values towards all cores. %within the global-collapsing-core scale.
As $\sin\omega$ is a measure for the magnetic field's efficiency to oppose gravity, its small values naturally are consistent with a gravity-induced alignment where gravity can entrain the B-field and form channels where material is preferentially
moving and accreting along the B-field lines towards a central denser region. 
This is precisely the                                  
%schematic scenario proposed in \citet{busquet20} and 
scenario put forward in \citet{wang20} for the G$33.92+0.11$ hub-filament system based on a detailed analysis of 
local correlations between
magnetic field, gravity, and velocity gradients, and it is identical with the cartoon in \citet{busquet20} illustrating 
the alignment transition discussed in \citet{pillai20} at visual extinctions $A_v \gtrsim 21$.
In summary, the quantitative analysis here based on $\sin\omega$, $|\mathbf{g}_{\rm{loc}}|$, and the B-field's stabilizing role 
(Figure \ref{figure_stability_criterion}) 
is suggestive of an {\it emerging picture where the detected dust lanes form fundamental accretion 
regions and channels where gravity and B-field are prevailingly aligned with these structures.
The dust lanes are likely gravitationally driven, seen in an underlying gravitational skeleton (Figure \ref{figure_grav_stream}) and magnetically stabilized.}
Such structures have been found in various simulations investigating the evolution of molecular clouds and 
accretion onto denser structures \citep{koertgen15,gomez18,li18}.

\subsection{Self-Similarity and Multi-Scale Embedded Collapse: Imprint onto B-field Morphology and Gravitational Field} 
\label{discussion_multi_scale_collapse}

The previous section has described four distinct scales with four distinct physical regimes where the role
of the magnetic field is quantified with various novel measures. 
Identifying these distinct scales is the result of a series of observations with successively higher resolutions. 
With these higher interferometric resolutions more and more of the surrounding diffuser emission is 
stepwise resolved out. Hence, the gradually higher resolutions are probing more and more of the inner
and denser regions. Patching this together leads to a picture from a pc-scale encompassing envelope 
to mpc-scale core-accreting dust lanes. 

\begin{figure}
\includegraphics[width=20cm, angle=90]{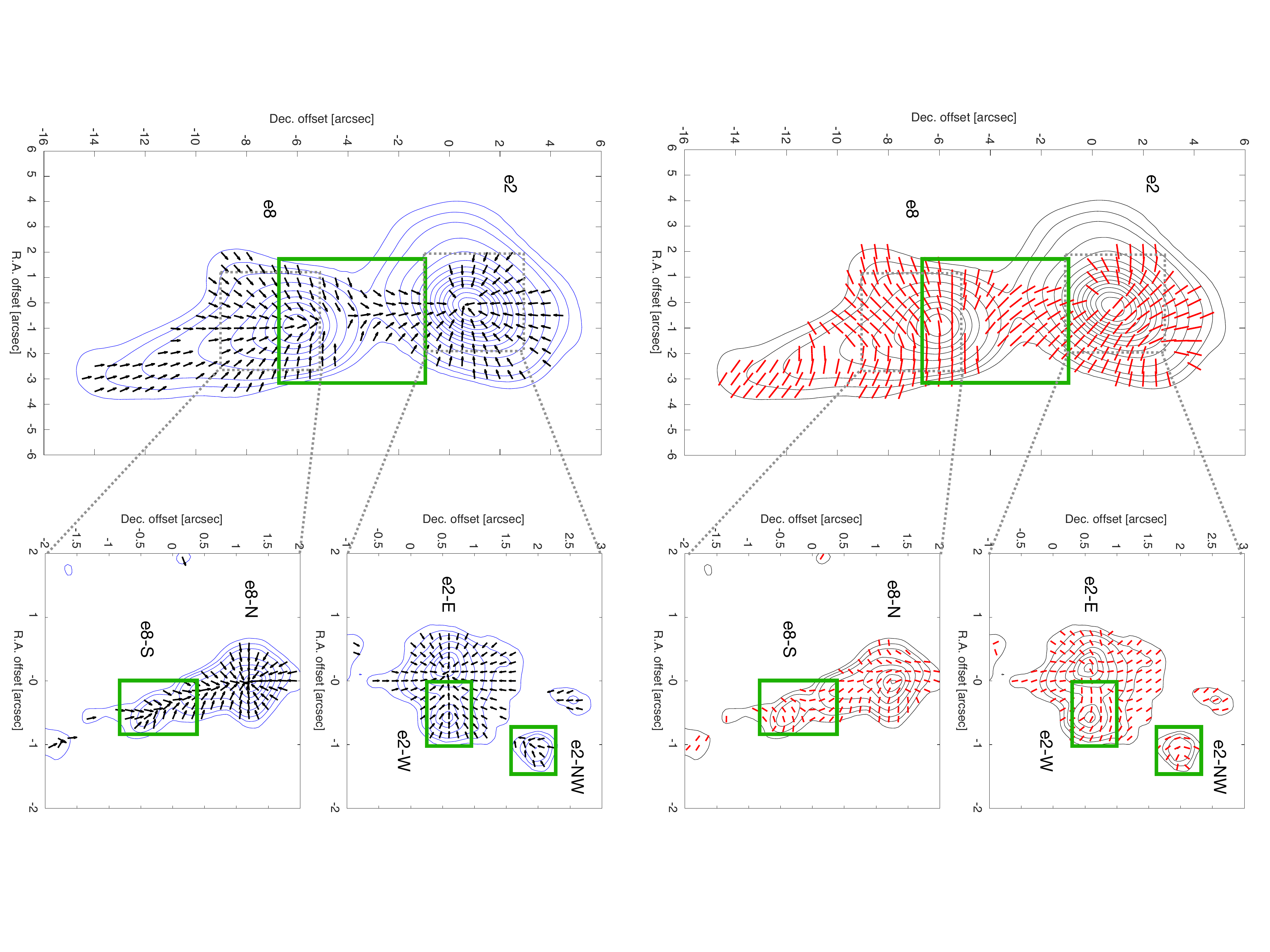}
\caption{\scriptsize Self-similar structures on two different scales (envelope-to-core scale (section 
\ref{discussion_envelope_to_core}; left column) and 
local-collapsing-core scale (section \ref{discussion_local_collapsing_core}; right column) 
in their detected magnetic field morphologies (red segments, top row) and their
gravitational vector fields (black arrows, bottom row).
Green boxes mark the self-similar structures. 
On the envelope-to-core scale between e8 and e2, the gravitational field starts to be deflected in the north of e8, 
gradually pointing more towards e2. The more massive e2 core mostly displays an azimuthally symmetric gravitational 
field.
This systematic deflection towards the more massive local center is seen again on smaller scales when zooming in between the cores
e2-W and e2-E, e2-NW and e2-E/e2-W, and e8-S and e8-N.
The B-field morphology shows a characteristic feature where the field segments are bent away from the less massive
core and straightened towards the more massive local center.
These recurring structures point at a multi-scale collapse-within-collapse scenario which leaves its imprints both in the B-field and the gravitational vector field.}
\label{figure_self_similar_structures} 
\end{figure}

Here, we emphasize one particular aspect in this picture, namely the {\it repeating structures in the magnetic 
field morphology and the gravitational vector field across different scales.} This is most evident when 
comparing the envelope-to-core scale and the local-collapsing-core scale  
as highlighted in Figure \ref{figure_self_similar_structures}.
The W51 e2/e8 complex has its dominating mass centered on e2. As a consequence, on the envelope-to-core
scale, e2 displays the beginning of a global collapse (i.e., a collapse of the entire e2 core as an entity) 
with clearly bent field lines in 
its northwestern and southeastern zones. Differently, the less massive e8 core shows the beginning
of pulled-in field lines on only one end, namely in its south. At the northern end, the field lines appear to be 
{\it bending away}, from straight east-west-oriented lines around offset Dec. $\sim -5\arcsec$, and {\it turning to be more
redirected towards the center of e2} (Figure \ref{figure_self_similar_structures}, top left panel).
On a smaller scale, {\it inside the global collapse of e2 and e8}, the magnetic field morphologies in e8-S, e2-W, and e2-NW
show a similar structure, i.e., bent and likely gravitationally pulled-in field lines on one end (southern end in e8-S, western 
end in e2-W, western end in e2-NW) and straightened, opened up, and redirected lines on the other end
(northern end in e8-S, eastern side in e2-W, eastern/south-eastern side in e2-NW; top right panels in Figure \ref{figure_self_similar_structures}). 
An analogous picture is seen in the gravitational vector field (bottom panels in Figure \ref{figure_self_similar_structures}).
Section \ref{analysis_local_gravity} has described the trend of the local gravitational field converging towards
the local peak emission on one side of a core while the other side is revealing local gravity to be pointing away
towards the neighboring dominating mass concentration. This is seen for e8-S towards e8-N, e2-W towards e2-E, 
and e2-NW towards e2-E/e2-W. An identical pattern in the gravitational vector field, on a larger scale, 
is observed between e8 (as an entity) and the northern more massive e2 (as an entity). 
In short, these observations show {\it self-similar structures in magnetic field and gravitational vector field}
on scales of 0.5~pc and 0.05~pc. {\it This local collapse happening simultaneously with the global collapse 
points at a collapse-within-collapse scenario.}

Recent numerical works by \citet{gomez18, vazquez19} suggest a multi-scale, global hierarchical collapse scenario,
described as a flow regime of collapses within collapses.
This matches very well our multi-scale observational findings of B-field morphologies and gravitational 
field. In the global hierarchical collapse by \citet{vazquez19}, all scales accrete material from their parent structures 
and filamentary accretion is a natural consequence of gravitational collapse. 

Finally, we note that zooming out to an even larger scale -- encompassing the W51 e2/e8 complex and the 
W51 North complex -- a self-similar structure is also present \citep[Figure 1 in][]{koch12b}. An analysis of the magnetic 
field tension-to-gravity force ratio $\Sigma_{\rm{B}}$ shows the largest values above one in between the 
two complexes with ratios systematically decreasing and falling below one towards both complexes.
The same trend in $\Sigma_{\rm{B}}$ is then also seen with higher-resolution observations for W51e2 and e8
and further within e2 \citep{koch12b}.
Therefore, we argue that all together the W51 region is displaying self-similar multi-scale collapse features across three 
different scales.

\section{Summary and Conclusion}

We are presenting ALMA continuum polarization observations towards the high-mass star-forming regions 
W51 e2 and e8 in Band 6 (around 230~GHz; $\lambda\sim$1.3~mm) with a resolution of $0\farcs1$ which corresponds to about 540~au.
Together with earlier observations we are proposing a multi-scale synergetic scenario for the different roles of 
the magnetic field from the large-scale 0.5~pc envelope to core-connecting networks of dust lanes
with widths of a few 100 au. 
Our main results are summarized in the following.

\begin{enumerate}

\item{{\it Polarization Detection.} 
Polarized emission at a resolution of about 540 au is clearly detected in W51 e2 and e8. 
Polarization percentages range from about 0.1\% to about 15\%, with averages around 3\%. 
A typical anti-correlation between polarization percentage and Stokes $I$ is observed, though with 
a large scatter and shallower slopes (around $-0.8$) than in coarser observations in the same source.
}

\item{{\it New Structural Features.}
These latest observations mark a departure from spherical and elongated structures seen in earlier observations. 
A connecting network of filamentary dust lanes is resolved in continuum with widths
around 540~au to 2,000~au.
 These dust lanes are
located in the periphery of cores and in between cores. The cores do not appear to fragment further. 
Magnetic field structures are resolved in dust lanes and extensions. A transition from perpendicular field lines
in the surrounding diffuser region to field lines aligning with the denser central spine in the connecting extension 
between cores is observed in e8. 
Three sectors in the inner $0\farcs5$ region around e2-E display nearly straight field lines that are rotated by $90^{\circ}$ 
from one sector to the next. These lines abruptly change orientation towards the innermost center, 
forming a northwest-southeast aligned structure, possibly hinting  
the presence of a small disk close to where a small-scale collimated SiO outflow is seen.
}

\item{{\it Gravitational Field and Magnetic Field Morphology.}
The magnetic field orientations are prevailingly aligned parallel to the dust lanes.
At the same time, the gravitational vector field 
%(specifying the local direction of gravity at every location 
%where a magnetic field orientation is measured) 
is typically also along the dust lanes, 
while additionally hinting a gravitational skeleton that defines ridges which likely form the underlying 
structures of the dust lanes. 
Hence, B-field and gravitational field show a very similar overall configuration.
This implies that gravitational pull is mostly directed along B-field lines.
A noticeably distinct feature in the gravitational field is the converging vector field on one side of a
core with the gravitational vector field on the other side of the core being gradually redirected towards 
the neighboring more massive core. 
}

\item{{\it Local Collapse Criterion and B-field-Induced Stability.}
Utilizing the detected magneto-gravitational field configuration, a local stability and collapse criterion is derived. 
This criterion is based on the measurable magnitude of local gravity and the measurable angle 
between the local orientations of gravity and magnetic field. The criterion sets a limit to the maximum
local magnetic field strength that can be overcome with an observed magneto-gravitational field configuration,
or in other words, it defines the necessary minimum field strength to prevent a local collapse.
The resulting field strengths suggest that the filamentary dust lanes and extensions are stable, possibly forming 
a fundamental structure in the accretion of material towards central sources and disks. 
The magnetic field can stabilize the dust lanes both against a local collapse and also against external pressure.
%This might point to a scenario for massive star formation without the need of large accretion disks.
}

\item{{\it Synergetic Picture and Multi-Scale Collapse.}
Combining resolutions starting from pc-scale to the latest mpc-scale, four
distinct scales can be identified in the W51 region: envelope-to-core scale, global-collapsing-core scale, 
local-collapsing-core scale, accreting-dust-lane scale. 
The roles of the magnetic field differ with these scales.
Various diagnostic tools are summarized to quantify the different roles. 
Repeating structures in the B-field morphology and the gravitational field are seen on different scales. 
These self-similar structures suggest a multi-scale collapse-within-collapse scenario. This starts from scales of several pc
until the eventual emergence of a network of stable core-connecting and accreting dust lanes on mpc scale 
where gravity and magnetic field are all aligned with the dust lanes, suggestive of a gravity-entrained alignment.
Hence, the dust lanes are likely gravitationally driven and magnetically stabilized.
}

\end{enumerate}

\section*{Appendix: Polarization Properties}

\begin{figure}
\includegraphics[width=9cm]{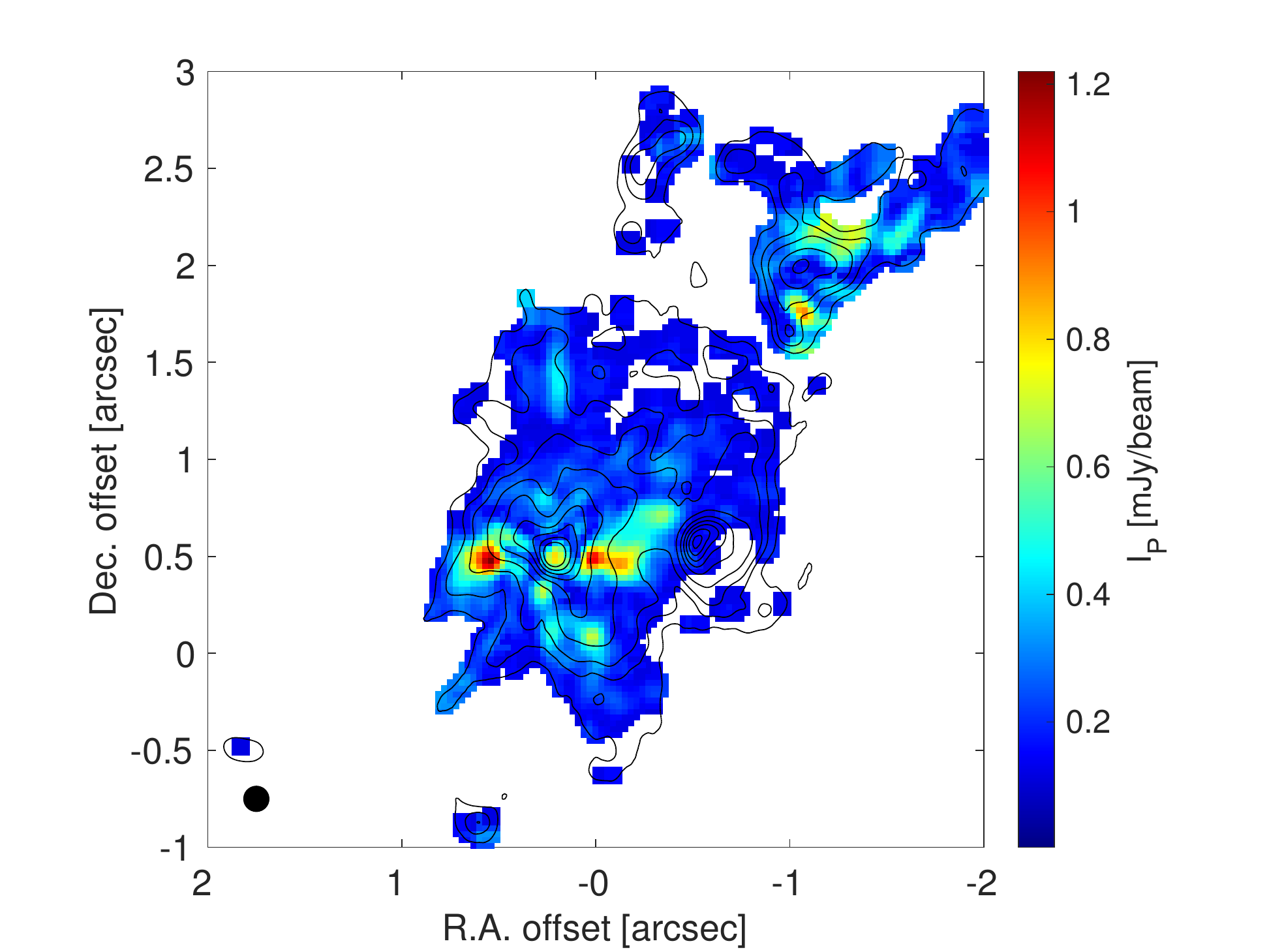}
\includegraphics[width=9cm]{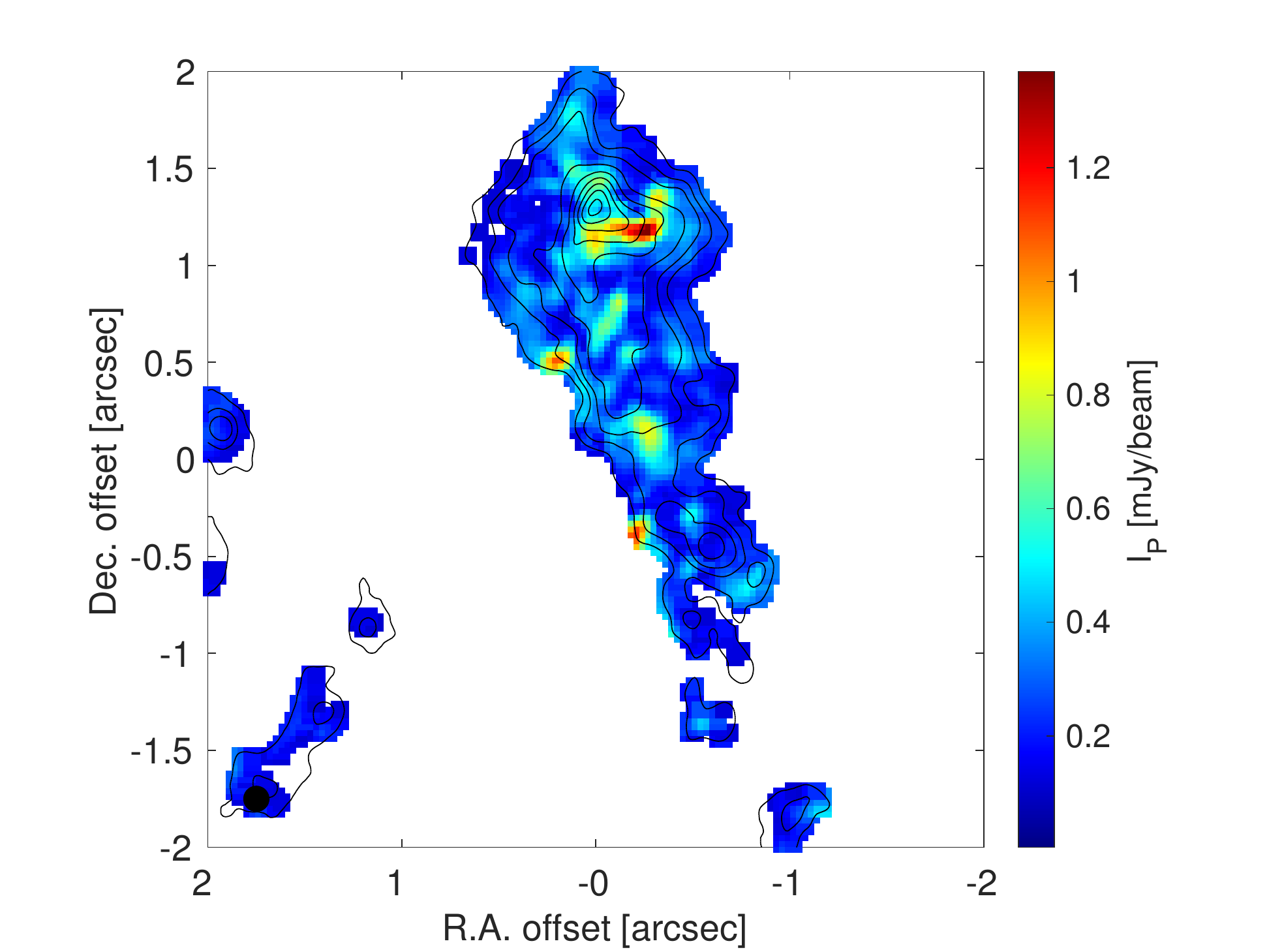}
\includegraphics[width=9cm]{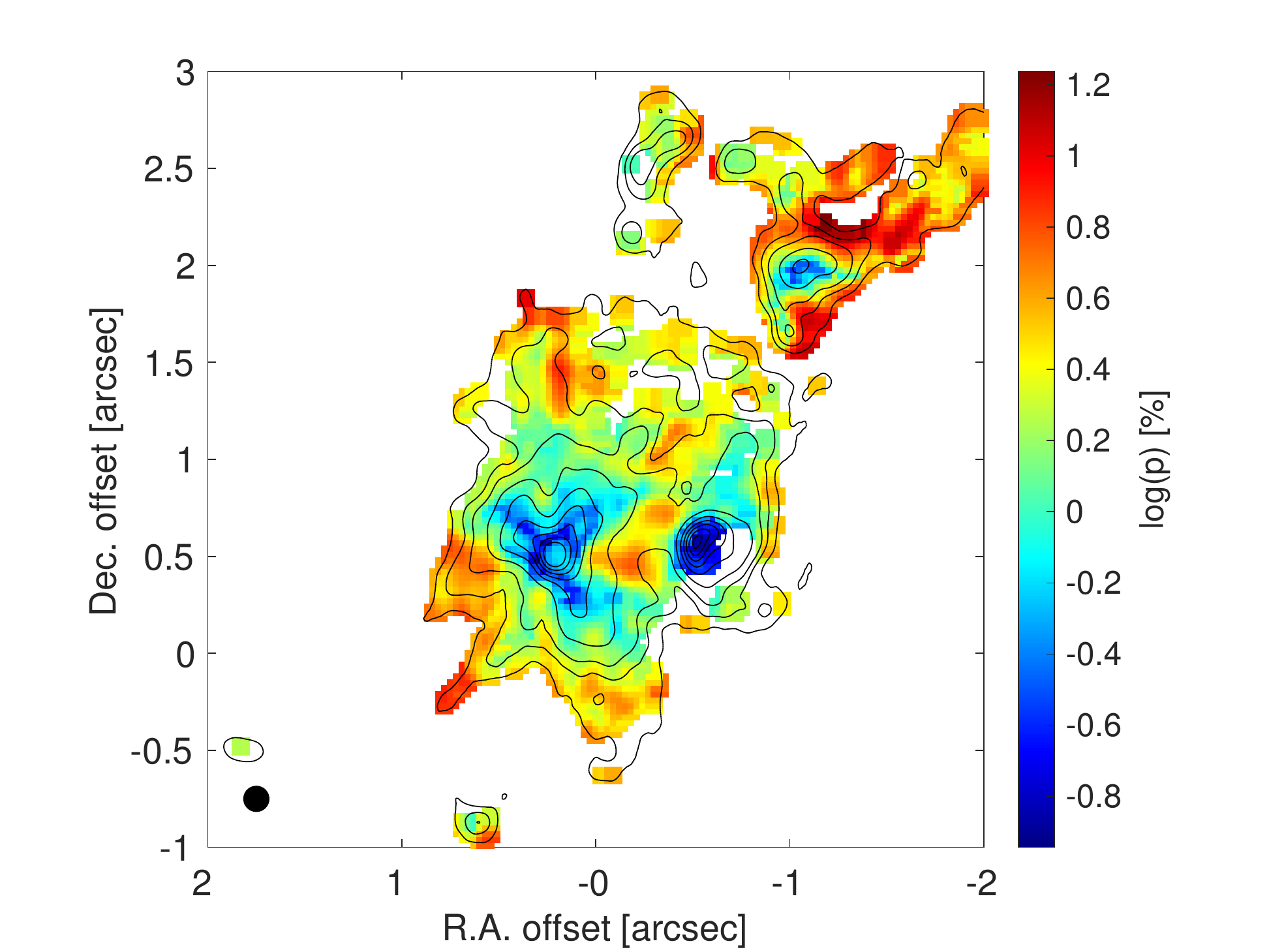}
\includegraphics[width=9cm]{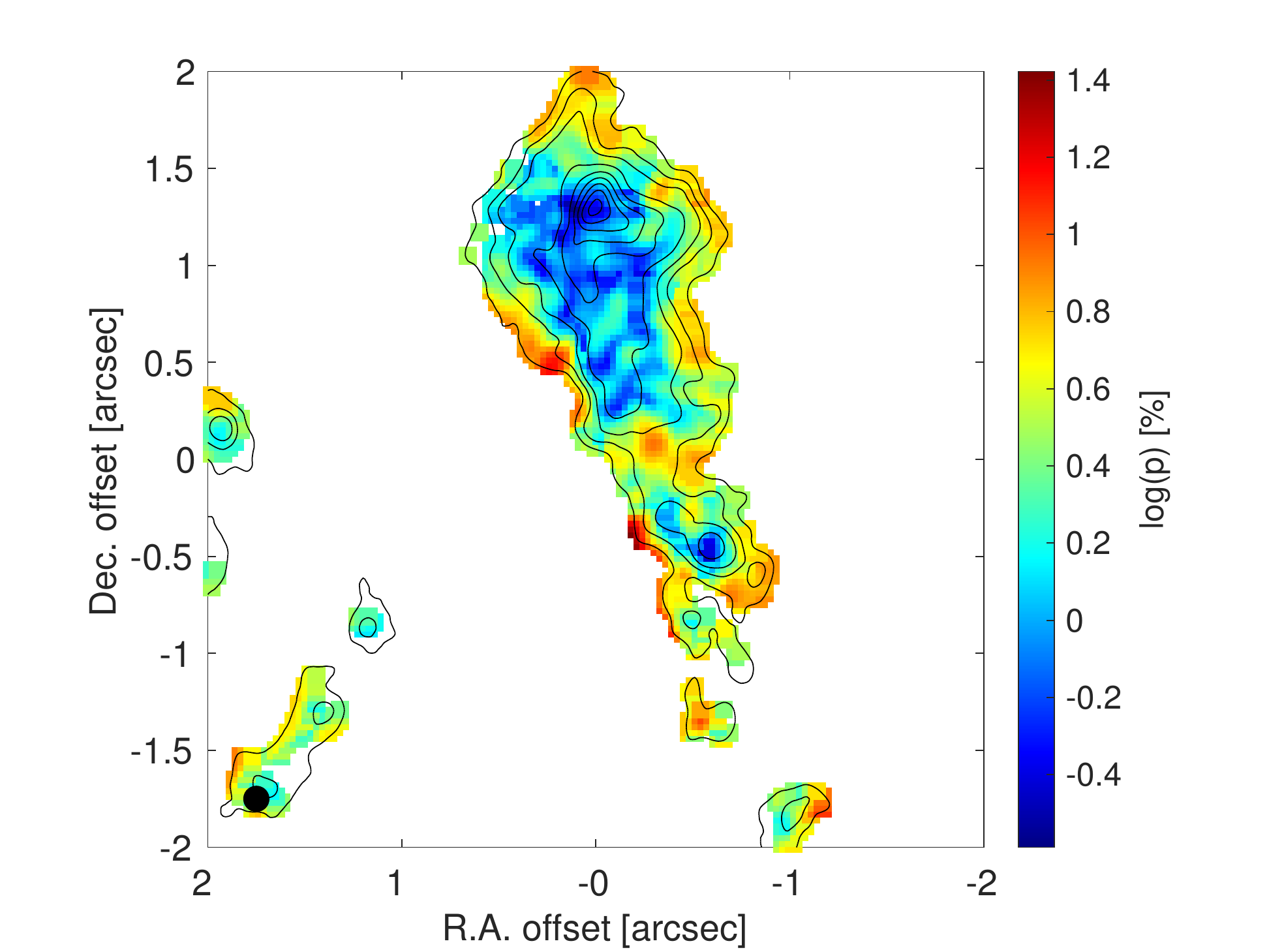}
\includegraphics[width=9cm]{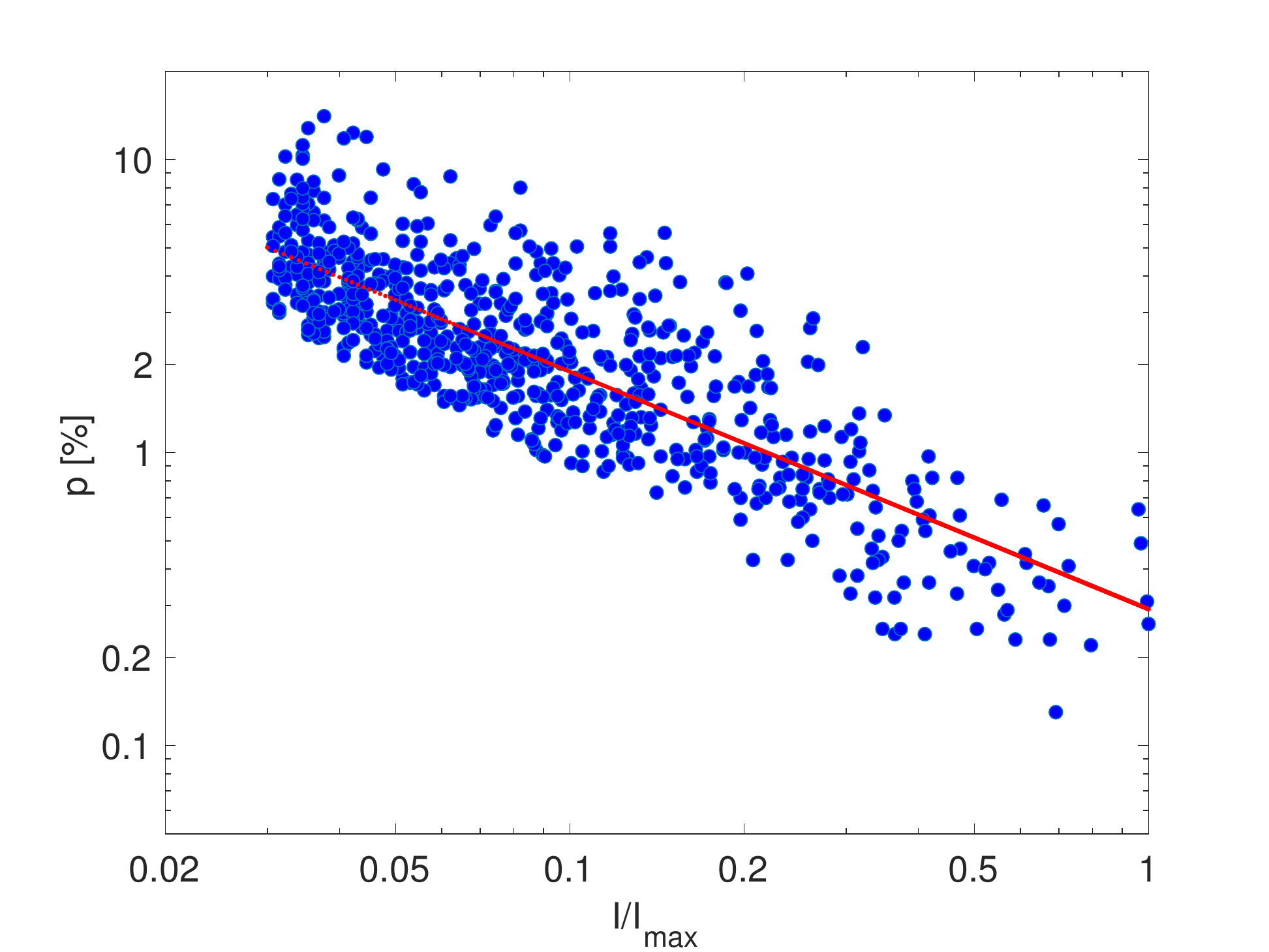}
\includegraphics[width=9cm]{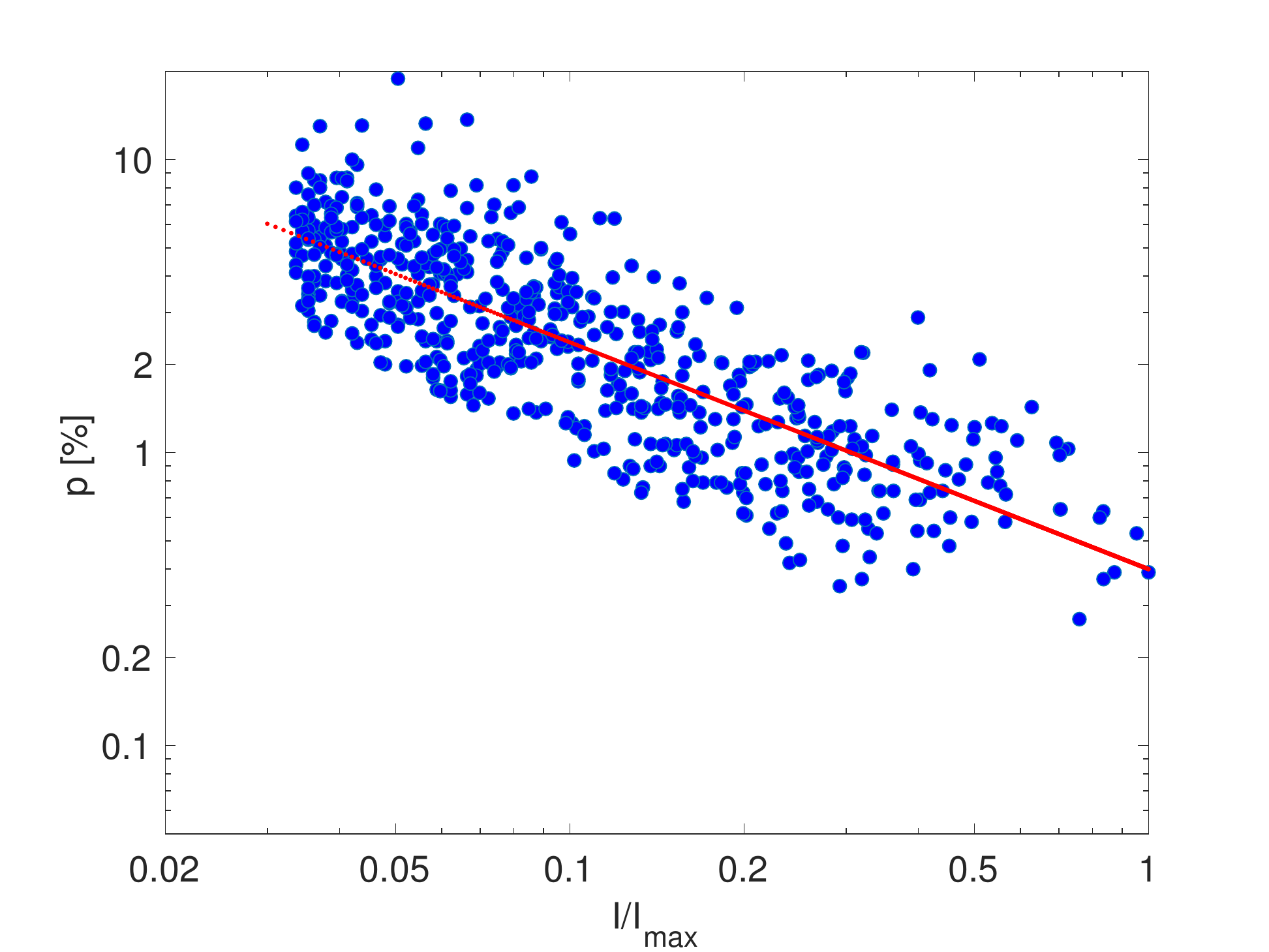}
\caption{\footnotesize Top to bottom panels: polarized emission $I_p$, polarization percentage $p$, and polarization percentage
versus Stokes $I$, normalized by $I_{\rm{max}}$, for W51 e2 (left column) and W51 e8 (right column). 
The red solid lines in the bottom panels are the best-fit power laws with indices $-0.81$ (e2) and $-0.77$ (e8).
The maps display overgridded data for a sharper visualization. The data in the bottom panels are extracted 
from maps gridded to only half of the synthesized beam resolution.}
\label{figure_polarization} 
\end{figure}

\begin{figure}
\includegraphics[width=9cm]{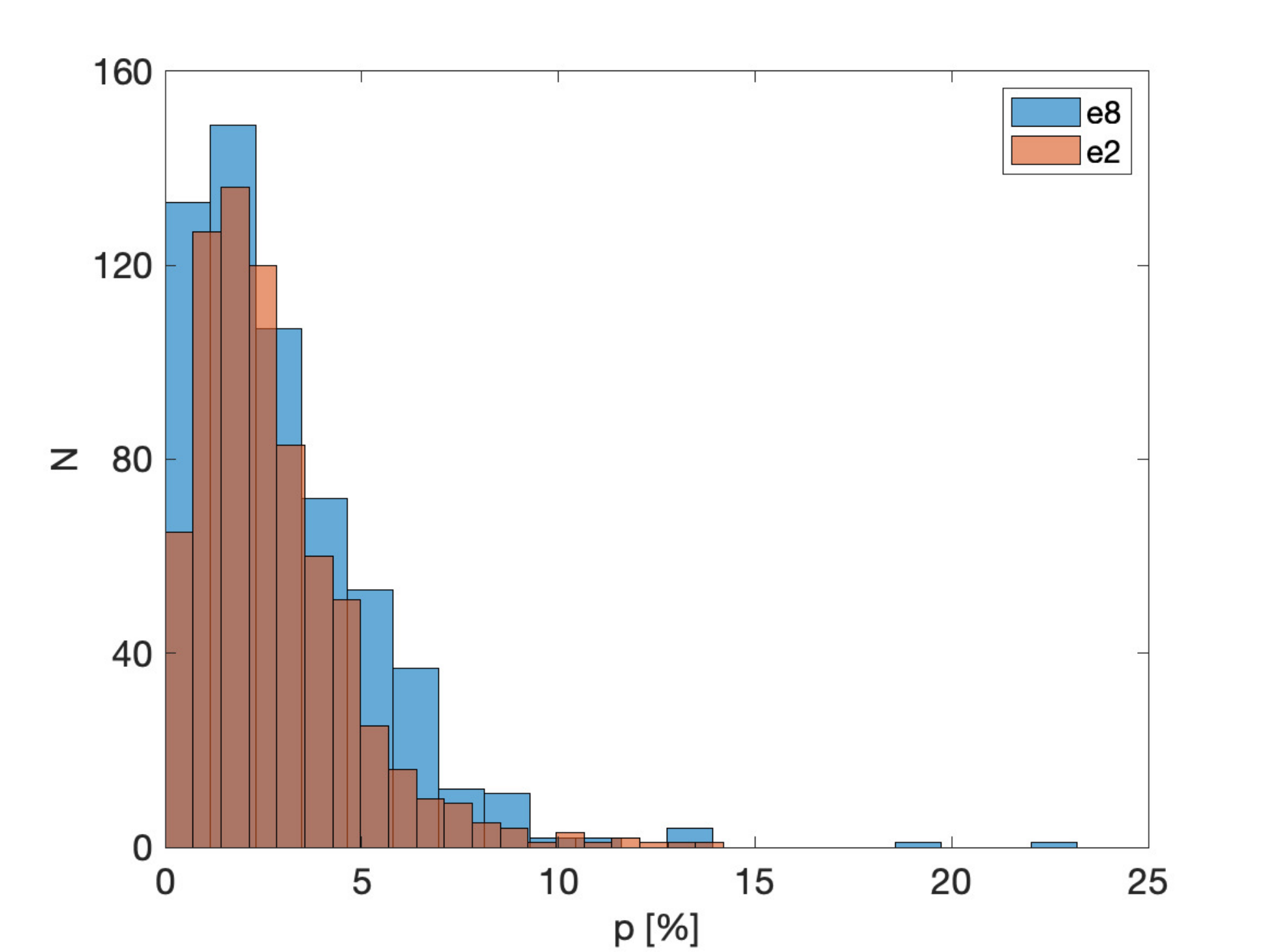}
\caption{Histograms of polarization percentages for W51 e2 and e8.
Averages, median values, and standard deviations are
2.8\%, 2.3\%, and 2.0\% for e2, and 3.1\%, 2.4\%, and 2.5\% for e8. 
%3.1% and 2.2%, 2.5% and 2.0%, and 2.3% and 2.3% for e2, e8, and North. 
Maximum and minimum polarizations are 
14\% and 0.13\% for e2 and 23\% and 0.27\% for e8.
}
\label{figure_hist_pol_perc} 
\end{figure}

The basic polarization characteristics are displayed in the Figures \ref{figure_polarization} and
\ref{figure_hist_pol_perc}.  
Typical trends are observed, namely a generally growing polarized signal $I_p$ towards the dust continuum
peaks with a generally decreasing polarization percentage $p=I_p/I$. We note that the peaks in $I_p$ are offset
from the Stokes $I$ peaks, both in e2 and e8, and e2 appears to have two peaks with one to the east and one to the west 
(Figure \ref{figure_polarization}). 
While the common anti-correlation $p$ versus $I$ is seen, the large vertical scatter of about one order
of magnitude indicates that there are likely more complicated physical processes ongoing which are not captured
with this simple anti-correlation. 
The slopes for the best-fit power laws are slightly different for e2 ($-0.81$) and e8 ($-0.77$). 
They appear to be shallower than the ones derived from the coarser observations at the same frequency 
($\theta \sim 0\farcs26$, 230 GHz; \citet{koch18}) 
which yielded $-1.02$ and $-0.84$ for e2 and e8, respectively. 
Polarization percentages with maxima, minima, and standard deviations (Figure \ref{figure_hist_pol_perc})
are similar to the values from the coarser observations.

%%%%%%%%%%%%%%%%%%%%%%%%%%%%%%%%%%%%%%%%%%%%%%%%%%%%%%%%%%%%%
%%%%%%%%%%%%%%%%%%%%%%%%%%%%%%%%%%%%%%%%%%%%%%%%%%%%%%%%%%%%%

%\begin{figure}
%\includegraphics[width=10cm]{figures/w51_01_e2_B.eps}
%\includegraphics[width=10cm]{figures/w51_01_e8_B.eps}
%\caption{Magnetic field orientations (red segments) observed with ALMA with a resolution of 
%$0\farcs1$ ($\sim 2.6$ ~mpc or $\sim 540$~au) around 210~GHz in Band 6 for W51 e2 (left panel) 
%and e8 (right panel) overlaid on dust continuum contours (black). 
%Magnetic field segments are displayed at half of the synthesized beam resolution (black ellipse in the lower
%left corner). 
%}
%\label{figure_B} 
%\end{figure}

%\begin{figure}
%\includegraphics[width=14cm]{figures/alma_cont3p_e2.eps}
%\includegraphics[width=12cm]{figures/alma_cont_zoom.eps}
%\includegraphics[width=14cm]{figures/alma_cont3p_e8.eps}
%%\includegraphics[width=12cm]{figures/alma_cont_zoom.eps}
%\caption{\small Magnetic field orientations (red segments) observed with ALMA with a resolution of 
%$0\farcs1$ ($\sim 2.6$ ~mpc or $\sim 540$~au) around 210~GHz in Band 6 for W51 e2 (left panel) 
%and e8 (right panel) overlaid on dust continuum contours (black). 
%Magnetic field segments are displayed at half of the synthesized beam resolution (black ellipse in the lower
%left corner). 
%}
%\label{figure_B} 
%\end{figure}

% \begin{figure}
%%\includegraphics[width=13cm]{w51_e2_01_grav_stream.eps}
%%\includegraphics[width=9cm]{w51_e2_01_grav_stream.eps}
%\includegraphics[width=9cm]{grav_try_1.eps}
%\includegraphics[width=9cm]{grav_try_4.eps}
%%\includegraphics[width=13cm]{grav_try_1.eps}
%\includegraphics[width=9cm]{grav_try_2.eps}
%\includegraphics[width=9cm]{grav_try_3.eps}
%%\includegraphics[width=13cm]{w51_01_e8_loc_grav.eps}
%\caption{TRY}
%\label{figure_loc_grav_stream} 
%\end{figure}

% \begin{figure}
%\includegraphics[width=9cm]{w51_01_e2_sin_omega.eps}
%\includegraphics[width=9cm]{w51_01_e8_sin_omega.eps}
%\includegraphics[width=9cm]{w51_01_e2_g_loc.eps}
%\includegraphics[width=9cm]{w51_01_e8_g_loc.eps}
%\includegraphics[width=9cm]{w51_01_e2_B.eps}
%\includegraphics[width=9cm]{w51_01_e8_B.eps}
%\caption{Local magnetic field strength maps (bottom row) derived from the stability and collapse criterion given in equation (\ref{eq_collapse_criterion}).  
%Separately displayed are the measurable input parameters $\sin\omega$ (top row) and $|\mathbf{g}_{\rm{loc}}|$ (middle row) 
%for W51 e2 (left column) and e8 (right column).
%}
%\label{figure_stability_criterion} 
%\end{figure}

%\begin{figure}

%\begin{figure}

%\begin{sidewaysfigure}
%%\includegraphics[width=22cm]{figures/artist_2.PNG}
%%\includegraphics[width=22cm]{figures/lic-w51e2-all.pdf}
%\includegraphics[trim=1.5cm 7.5cm 8.5cm 1.5cm, clip=true,width=22cm]{w51-all-lic_all_scales.pdf}
%\caption{\small Visualization of the evolving roles of the magnetic field over the scales presented in Figures 
%\ref{figure_synergy_1} and \ref{figure_synergy_2}.
%Visualized are streamlines based on the detected magnetic field morphologies in Figure \ref{figure_synergy_1}
%using a line
%convolution algorithm for vector fields (e.g., \citet{cabral93}).
%From left to right are envelope-to-core scale (first and second), 
%global-collapsing-core scale (third),
%local-collapsing-core scale (forth), and {\bf accreting-dust-lane} scale (fifth).
%No visualization is shown in black patches with too sparse sampling in the original data
%which does not allow to construct a connected convolution. 
%}
%\label{figure_visualization} 
%\end{sidewaysfigure}

%\begin{figure}
%\includegraphics[width=18cm]{synergetic_picture_4_scales_2.pdf}
%\caption{Synergetic picture of the evolving roles of the magnetic field over the scales presented in Figures 
%\ref{figure_synergy_1} and \ref{figure_synergy_2}.
%}
%\label{figure_synergy_3} 
%\end{figure}

% figure appendix for polarization

% \begin{figure}
%\includegraphics[width=9cm]{w51_01_e2_pol_abs.eps}
%\includegraphics[width=9cm]{w51_01_e8_pol_abs.eps}
%\includegraphics[width=9cm]{w51_01_e2_pol_perc.eps}
%\includegraphics[width=9cm]{w51_01_e8_pol_perc.eps}
%\includegraphics[width=9cm]{w51_01_e2_pol_perc_vs_I.eps}
%\includegraphics[width=9cm]{w51_01_e8_pol_perc_vs_I.eps}
%\caption{\footnotesize Top to bottom panels: polarized emission $I_p$, polarization percentage $p$, and polarization percentage
%versus Stokes $I$, normalized by $I_{\rm{max}}$, for W51 e2 (left column) and W51 e8 (right column). 
%The red solid lines in the bottom panels are the best-fit power laws with indices $-0.81$ (e2) and $-0.77$ (e8).
%The maps display overgridded data for a sharper visualization. The data in the bottom panels are extracted 
%from maps gridded to only half of the synthesized beam resolution.}
%\label{figure_polarization} 
%\end{figure}

{\it Facilities:} \facility{ALMA}.

\acknowledgments{
The authors thank the referee for valuable comments which
led to further insight in this work.
The authors thank C. Goddi for making available the outflow data for W51 e2 and e8.
This paper makes use of the following ALMA data: 
ADS/JAO.ALMA\#2013.1.00994.S, ADS/JAO.ALMA\#2016.1.01484.S, and 
ADS/JAO.ALMA\#2017.1.01242.S.
 ALMA is a partnership of ESO
(representing its member states), NSF (USA), and NINS (Japan),
together with NRC (Canada), MoST, and ASIAA (Taiwan), and
KASI (Republic of Korea), in cooperation with the Republic of
Chile. The Joint ALMA Observatory is operated by ESO, AUI/
NRAO, and NAOJ.
ADC acknowledges the support from the Royal Society via a University Research Fellowship (URF/R1/191609). 
Y-WT is supported by the Ministry of Science and Technology (MoST) in Taiwan through grant
MOST 110-2112-M-001-035 -
and MOST 108-2112-M-001-004 -MY2.
PMK acknowledges support from
MoST 108-2112-M-001-012, MoST 109-2112-M-001-022, 
MoST 110-2112-M-001-057, and from an Academia Sinica Career Development Award.
}
%
%
% REFERENCES

\bibliographystyle{apj}                       %% AASTeX
\bibliography{w51}
\end{document}